\documentclass[aps,pra,twocolumn,floatfix,nofootinbib]{revtex4-1}
\usepackage{amsthm}
\usepackage{amsmath}
\usepackage{latexsym}
\usepackage{amsfonts}
\usepackage{amssymb}

\usepackage[dvipsnames]{xcolor}
\usepackage{bbm,dsfont}
\usepackage{graphicx}
\usepackage{mathrsfs}
\usepackage{mathbbol}
\usepackage{hyperref}
\usepackage{enumerate}
\usepackage{MnSymbol}
\usepackage{bbding}
\usepackage{float}
\usepackage{soul}
\usepackage{ulem}
\usepackage{physics}
\newcommand{\Ket}[1]{\vert #1 \rrangle}
\newcommand{\Bra}[1]{\llangle #1\vert}
\usepackage[T1]{fontenc}
\usepackage{currvita}

\begin{document}

\title{Synchronization and coalescence in a dissipative two-qubit system}

\author{Albert Cabot}
\affiliation{IFISC (UIB-CSIC), Instituto de F\'isica Interdisciplinar y 
Sistemas 
Complejos, Palma de Mallorca, Spain}

\author{Gian Luca Giorgi}
\affiliation{IFISC (UIB-CSIC), Instituto de F\'isica Interdisciplinar y 
Sistemas 
Complejos, Palma de Mallorca, Spain}

\author{Roberta Zambrini}
\affiliation{IFISC (UIB-CSIC), Instituto de F\'isica Interdisciplinar y 
Sistemas 
Complejos, Palma de Mallorca, Spain}

\begin{abstract}

The possibility for detuned spins to display synchronous oscillations in local observables is analyzed in the presence of collective dissipation and  incoherent pumping. We show that there exist two distinct mechanisms that can give rise to synchronization, that is, non-degenerate subradiance and coalescence. The former, known as transient synchronization, is here generalized in the presence of pumping. It is due to  long-lasting coherences leading to a progressive frequency selection.  In the same set-up, even if under different conditions, coalescence and exceptional points are found  which can lead to regimes where  a single oscillation frequency is present in the relevant quantities. Still, we show that synchronization can be established only after steady phase-locking occurs. Distinctive spectral features of synchronization by these two different mechanisms are reported for two-time correlations.
\end{abstract}

\maketitle

\section{Introduction}
\label{sec1}

Open quantum systems  exhibit features  beyond dissipation of energy and decoherence that can not generally be found in the absence of losses \cite{Breuer}. An example studied in the last decade is spontaneous synchronization emerging among different  interacting quantum systems, reaching a synchronized dynamics determined by the coupling to some external environments \cite{SyncRev1}. Different approaches have been proposed to define and describe this phenomenon in the quantum regime, also considering a variety of systems such as
harmonic oscillators \cite{Giorgi1,manzano,cabot_npj}, spins \cite{praspins,3qubits}, biological \cite{olaya} or optomechanical \cite{mari,marquardt,Cabot_NJP} systems, quantum van der Pol oscillators \cite{lee,lee2,walter2,walter,tilley1}  or micromasers \cite{tilley2}, also exploring the effects for different system-bath configurations \cite{praprob,Bellomo,Cabot_PRL}. Synchronization signatures between mesoscopic ensembles of atomic systems have also been discussed in \cite{Holland1,Holland2} using a semiclassical approach.

In particular, quantum synchronization can be induced by dissipation when time-scale separation occurs between the modes governing the dynamics  \cite{SyncRev2},
due to the presence of a dominant collective excitation. 
Depending on the lifetime of this excitation, this synchronization can be  either observed  in a transient regime prior to thermalization, or found in the stationary dynamics in the presence of decoherence-free channels \cite{manzano,cabot_npj,dieter1,dieter2}. From a  mathematical point of view, when describing the open quantum system through a master equation, this dominant collective excitation emerges if one  eigenmode of the Liouvillian has a decay rate much smaller than any other eigenmode.  This analysis provides a clear criterion to predict transient synchronization, even if other scenarios can occur in more complex systems, as the recently reported band synchronization   \cite{Cabot_PRL}, where a bunch of weakly damped eigenmodes are almost degenerate. Then, synchronization is associated with the presence of a spectral gap that makes the long-time dynamics almost monochromatic. Different  measures can be used to characterize quantum synchronization  as reviewed in  \cite{SyncRev1}, including temporal correlations of local observables or  properties of the Liouvillian spectrum.

Another very interesting phenomenon displayed by open systems is the existence of spectral singularities, the so-called exceptional points (EPs) \cite{heiss}: in such points,  two or more eigenvalues, and their corresponding eigenvectors, simultaneously coalesce (i.e. one or more eigenvectors disappear) making the dynamics not diagonalizable. The presence of these singularities has been mainly studied, among other contexts, in the framework of $PT $-symmetric quantum mechanics \cite{bender} and non-Hermitian Hamiltonians \cite{El-Ganainy,feng,stefano,miri,ozdemir}, nontrivial  transmission and fluctuation spectra \cite{Cabot_EPL}, anomalous decay dynamics \cite{Cabot_EPL,Longhi1}, characterization of topological materials \cite{ghatak}, and enhanced sensing \cite{chen,stefano3}. The study of the dynamical behavior near  EPs has attracted interest especially  in integrated photonics \cite{peng,miao,stefano2,hodaei}, acoustics \cite{fleury,ding,shi}, and optomechanics \cite{lu,xu,verhagen}. 

A common feature shared by transient synchronization and eigenvalue coalescence is the reduction of the number of modes with different frequencies observed in the dynamics. The existence of common dynamical signatures, such as the presence of a single frequency in the temporal evolution of coupled systems, allows for the achievement of a synchronous dynamics in both cases. For instance, in Ref. \cite{Holland1},  the dynamics of two detuned atomic clouds interacting with a cavity mode and externally pumped was studied using a semiclassical approach. The identified  regime in which the system  displays only one frequency is indeed an example of synchronization by coalescence, as we will discuss here.  

The aim of this work is to make  a deep analysis and comparison between the synchronization dynamics emerging in both scenarios, that is when there is coalescence or when there is a weakly damped non-degenerate eigenmode, in order to establish their relation and distinctive signatures. Both phenomena can be displayed in a simple system of two spins interacting through a common bath. By means of an explicit diagonalization of the Liouvillian superoperator governing the dynamics, we will be able to fully characterize the regimes where (some of) the eigenmodes can coalesce and compare them with the synchronization diagram, which can be drawn either looking at temporal correlations between local observables or at the presence of a gap in the Liouvillian spectrum.

The emergence of both frequency- and phase-locking in transient synchronization has been shown to be due to frequency selection and long-lasting coherences between the ground and the slow eigenmode that emerges because of  non-degenerate subradiance   \cite{Bellomo}.
We will show that instead, in the presence of coalescence, a monochromatic oscillation is present from the beginning. Nevertheless, synchronization occurs after a transient anyway, as phase-locking emerges only when all (frequency-degenerate) eigenmodes but one have decayed out. 
Interestingly, coalescence is actually associated to the phenomenon of $degenerate$ super/subradiance, a connection unnoticed in the literature.
Furthermore, we will analyze the signatures of these two distinct  mechanisms of synchronization due to coalescence (degenerate subradiance)  and non-degenerate subradiance in the two-time correlation spectrum of the system.

The paper is organized as follows. In Sec. \ref{model} we present the model of an open system of two coupled qubits. In Sects. \ref{EPs} and  \ref{sync} we analyze the presence of EPs in the Liouvillian, and compare it with transient synchronization. The distinctive signatures of both phenomena in the correlation spectrum are analyzed in Sec. \ref{spectrum}, while in Sec. \ref{manyqubit} we discuss some relevant results in a more general context of many-qubit scenarios. Finally, in Sec. \ref{discussion} we discuss the relation of our findings with other works, and we present our conclusions. Some mathematical details and supplemental results are presented in four appendices \ref{appA}, \ref{appB}, \ref{appC}, \ref{appd}.

\section{The model}\label{model}

We consider a dissipative system of two qubits  described by the following Born-Markov master equation for their density matrix $\hat{\rho}$  ($\hbar=1$)
\begin{equation}\label{ME}
\dot{\hat{\rho}}=-i[\hat{H},\hat{\rho}]+2\gamma\mathcal{D}[\hat{L}]+w(\mathcal{D}[\hat{\sigma}_1^+]+\mathcal{D}[\hat{\sigma}_2^+]),
\end{equation}
where we have introduced dissipative superoperators in the Lindblad form \cite{Breuer} $\mathcal{D}[\hat{o}]=\hat{o}\hat{\rho}\hat{o}^\dagger-\hat{o}^\dagger\hat{o}\hat{\rho}/2-\hat{\rho}\hat{o}^\dagger\hat{o}/2$, the rising and lowering operators $\hat{\sigma}_j^\pm$ for spin $j=1,2$ are defined as usual from the Pauli matrices  $\hat{\sigma}_j^{x,y,z}$, and  $\hat{L}=(\hat{\sigma}_1^-+\hat{\sigma}_2^-)/\sqrt{2}$. The Hamiltonian part of this model reads as
\begin{equation}\label{Ham}
\hat{H}=\frac{\omega_1}{2}\hat{\sigma}_1^z+\frac{\omega_2}{2}\hat{\sigma}_2^z+s_{12}(\hat{\sigma}_1^-\hat{\sigma}_2^++\hat{\sigma}_1^+\hat{\sigma}^-_2).
\end{equation}
and describes two detuned spins with $\delta=\omega_1-\omega_2$, and central frequency $\omega_0=(\omega_1+\omega_2)/2$, which interact coherently through the exchange term with rate $s_{12}$. Notice that two types of incoherent processes are taken into consideration: the qubits dissipate collectively through $\hat{L}$ and with rate $2\gamma$, and a local incoherent pumping acts on each spin with rate $w$. 

The phenomenological model that we consider lies in the context of recent experiments in which a small number of two-level systems are found to interact and to display signatures of collective dissipation as subradiant and superradiant effects. The nature of the two-level systems and the origin of these interactions are diverse, as for instance: trapped atoms \cite{atoms2,atoms3} and ions \cite{ions1} interacting through waveguides or cavity modes, photon-mediated interactions between color centers in diamond \cite{diamond1}, 
superconducting  qubits with photon-mediated interaction in 1D lines \cite{Wallraff1} or in the bad cavity limit \cite{Wallraff2}. Some theoretical works dealing with collective dissipation, as in our work, analyze the coupling to a common cavity mode in the bad cavity limit \cite{Holland1},  the coupling to a common structured bath \cite{Fernando,Tudela} or to an effective one-dimensional bath  as in waveguides \cite{waveguideCB}, photonic nanostructures \cite{Asenjo} or microwave transmission lines \cite{Blais}.
Furthermore, tailored local incoherent processes such as the incoherent pumping can be realized addressing auxiliary energy levels of the spin system \cite{Cirac1,Holland1}. 

Finally, an important remark on the parameter values is that we consider them to  follow a hierarchy given by $\omega_0\gg \delta,\gamma,s_{12},w$ and $w,\delta,s_{12}\sim \gamma$, as it is a usual requirement for this kind of phenomenological models to have a microscopic origin \cite{Bellomo,Marco1}. It is also important to notice that, depending on the microscopic origin of the model, some mutual dependencies between the values of the parameters might exist, however, in the spirit of exploring the full model, we do not consider these particular constraints in this work, and we enable the parameters to vary independently of each other.

\section{Exceptional points in the Liouvillian}\label{EPs}

\begin{figure*}[t]
 \centering
 \includegraphics[width=2\columnwidth]{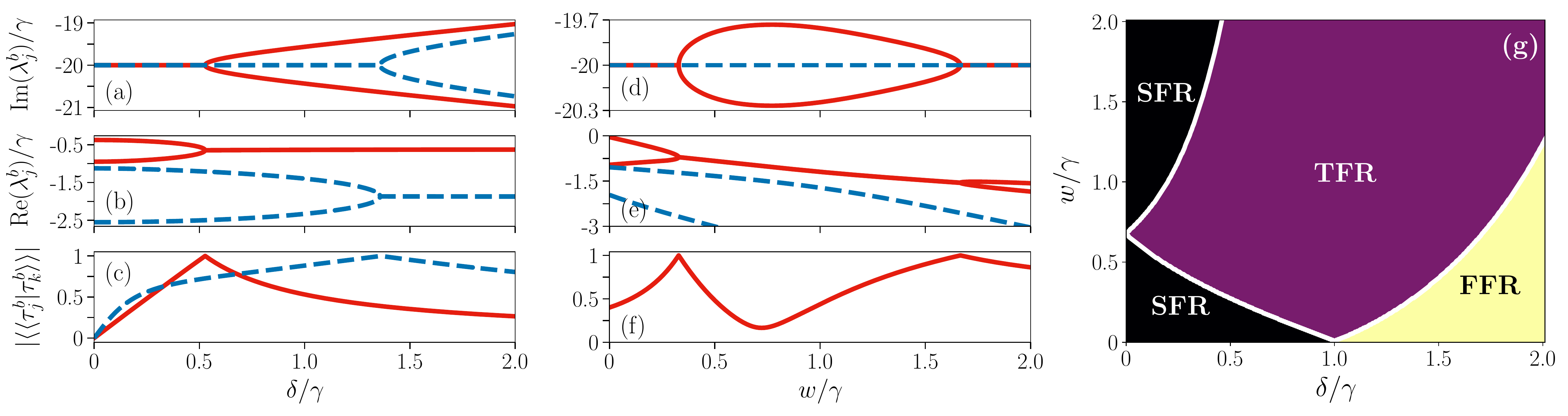}
 \caption{All panels: analysis of coalescence in $\mathcal{L}_b$ for $s_{12}/\gamma=0$ and $\omega_0/\gamma=20$. (a) Imaginary part of the eigenvalues (eigenfrequencies) varying $\delta/\gamma$, for $w/\gamma=0.25$. In solid red and dashed blue the two different pairs of eigenvalues that coalesce. (b) The real part of the corresponding eigenvalues (decay rates). (c) Product of the corresponding pair of eigenvectors that coalesce. (d)-(f) Same quantities as in (a)-(c) but fixing $\delta/\gamma=0.4$ 
 and varying  $w/\gamma$. Notice that the smallest (in absolute value) decay rate is not zero for $w/\gamma=0$ as it can be checked from Eq. (\ref{eigs_b1}). Here only a pair of eigenvectors coalesce (twice). (g) Diagram of eigenfrequencies. The white lines stand for second order EPs and separate the regions with different number of eigenfrequencies and decay rates. The lines $\delta/\gamma=0$ and $w/\gamma=0$ are not resolved in this plot, but analytical expressions are available (appendix \ref{appA}). In black, we have the SFR in which there is only one eigenfrequency ($-\omega_0$) and four decay rates. The two black regions are connected as $\delta/\gamma=0$ $w/\gamma=2/3$ is not an EP  [Eq. (\ref{eigs_b2})], while for $w/\gamma=0$ and $\delta/\gamma=1$ there are two second order EPs at the same point involving different decay rates [Eq. (\ref{eigs_b1})]. Moreover, there is an isolated EP for $w/\gamma=1$, $\delta/\gamma=0$ and $s_{12}/\gamma=0$ [see Eq. (\ref{eigs_b2})]. In purple, the TFR  where there are three eigenfrequencies and three decay rates. In yellow the FFR where four eigenfrequencies and two decay rates are found. The most prominent features are displayed in this range of detunings and pumping rates. } 
 \label{F123}
\end{figure*}

For our purposes, it is  convenient to describe the evolution of the two-spin density matrix within the Liouville formalism. Indeed, an isomorphism can be adopted which maps $\hat{\rho}$  into the $16$-dimensional vector $\Ket{\rho}$ and the Liouville super-operator into a $16\times 16$ matrix $\mathcal{L}$ \cite{Bellomo}.
 
The time evolution of the density matrix  can then be  rewritten as a vector equation $\Ket{\dot\rho}=\mathcal{L}\Ket{\rho}$. How to explicitly build $\mathcal{L}$ is  detailed in appendix \ref{appA}, where we generalize the results of \cite{Bellomo} to the case of incoherent driving.
 This matrix is block diagonal, $\mathcal{L}=\bigoplus_\mu \mathcal{L}_\mu$, with $\mu \in\{a,b,c,d,e\}$, the different blocks being related to the dynamics of different observables (in  appendix \ref{appA} we give the explicit expressions of such matrices). For instance, the dynamics of populations $\langle \hat{\sigma}^z_j\rangle$ is entirely described by $\mathcal{L}_a$, while the dynamics of coherences $\langle \hat{\sigma}^{x,y}_j\rangle$ by $\mathcal{L}_b$ and $\mathcal{L}_c=\mathcal{L}_b^*$. 
 Such a block structure is a direct consequence of  a symmetry on
the superoperator level, that is, the invariance of the Liouvillian 
under the action of the total-number-of-particles superoperator, and appears every time the (partial) secular approximation holds. Thus, it can be found in a very broad class of systems, as detailed in Ref. \cite{Marco2}.
 
In the study of synchronization we focus on the oscillatory dynamics of the coherences, and thus the analysis of the eigenspectrum of $\mathcal{L}_b$ and $\mathcal{L}_b^*$ yields the necessary information to assess the emergence of this phenomenon \cite{Bellomo,Giorgi1,Cabot_PRL}.
Within this formalism, 
the general solution of the master equation at time $t$ can be formally written as 
 \begin{equation}\label{rhot}
  \Ket{\rho(t)}=\sum_{\mu}\sum_{k} p_{0\, k}^{\mu}\, \Ket{ \tau^\mu_k} \,
\mathrm{e}^{\lambda_k^\mu t},
\end{equation}
where $\mu$ runs over the five blocks  of $\mathcal{L}$ and $k$ between $1$ and the dimension of the corresponding block. In Eq. (\ref{rhot}), we have introduced the right (left) eigenvectors of the Liouvillian $\Ket{
\tau^\mu_k}$ ($\Ket{\bar{\tau}^\mu_k}$), their respective eigenvalues $\lambda_k^\mu$, defined through $\mathcal{L} \Ket{\tau_k^\mu}=\lambda_k^\mu \Ket{\tau_k^\mu}$ ($\mathcal{L}^\dagger \Ket{\bar{\tau}_k^\mu}=\lambda_k^{\mu*} \Ket{\bar{\tau}_k^\mu}$) and the weight of the initial conditions  $p_{0\,
k}^{\mu}=\frac{\llangle \bar{\tau}^{\mu}_k \Ket{ \rho(0)}}{\llangle
\bar{\tau}^{\mu}_k\Ket{\tau^{\mu}_k}}$, where we use the Bra-Ket notation.  Notice that left and right eigenvectors form a biorthogonal basis: 
$\llangle \bar{\tau}_j^{\mu } \Ket{\tau_k^{\nu } }\propto\delta_{\mu \nu}\delta_{jk}$.

Being the system open, $\mathcal{L}_b$ ($\mathcal{L}$) is non-Hermitian, so it is actually    possible to have points in parameter space in which several eigenvalues and the corresponding eigenvectors coalesce, making the matrix non-diagonalizable \cite{Longhi1,NoriEPs}. These are the exceptional points (EPs) introduced in Sec. \ref{sec1}, whose order is defined as the number of eigenvalues and eigenvectors that coalesce. As anticipated, in this work we focus on the EPs occurring in $\mathcal{L}_{b(c)}$, as they are relevant for the emergence of synchronization. However, we notice that $\mathcal{L}_a$ is also able to display EPs as reported in appendix \ref{appA}. In Fig. \ref{F123} we analyze the presence of EPs in $\mathcal{L}_b$ for $s_{12}/\gamma=0$. We first show particular examples of the EPs by tuning $\delta/\gamma$ (a)-(c) and $w/\gamma$ in (d)-(f). Then, in panel (g), the overall picture is presented as a function of both detuning and pumping, showing the parameter regions where the Liouvillian displays from one to four frequencies: single-frequency regime (SFR), and similarly for three (TFR) and four (FFR). 

In Figs. \ref{F123}(a) and (d), we plot the imaginary part of the eigenvalues (eigenfrequencies), their real part (decay  rates) (b) and (e), and the absolute value of the product of the coalescing (normalized) eigenvectors $|\llangle\tau^b_j|\tau^b_k\rrangle|$ that is going to reach value one in the presence of coalescence, (c) and (f). Both EPs appearing in $\mathcal{L}_{b(c)}$ are second order; two eigenvalues become the same and the corresponding eigenvectors become linearly dependent, which makes the matrix non-diagonalizable. In (a)-(c), increasing $\delta/\gamma$ we observe a  common trend as the number of frequencies (decay rates) increases (decreases). While in this case the two EPs appear for different detunings, notice that for $w/\gamma=0$ these  arise for the same value $\delta=\gamma$ [Eq. (\ref{eigs_b1}) with $s_{12}=0$] where the term $V=\sqrt{\gamma^2-\delta^2}$ present in all eigenvalues vanishes. In this special case, $w/\gamma=0$, the emerging frequencies are degenerate and given by $\omega_0\pm\text{Im}(V)/2$. The physical intuition in this case is that the detuning needs to overcome the dissipation in order to induce the oscillatory behavior of the system, somehow analogously to an overdamped to underdamped transition, but keeping in mind that here $\omega_0/\gamma\gg1$.

While there was a common trend in the emergence of EPs for increasing detuning, the number of frequencies and the related appearance of EPs is more complex for increasing pumping. For small detuning (and still vanishing coupling  $s_{12}$) only one frequency is present into the system; then increasing it beyond a first EP we find a TFR and then again SFR. From \ref{F123}(d)  we also notice that is the same pair of eigenvectors that coalesce (twice). Furthermore, the pair of EPs disappears for vanishing detuning with the frequency separation (closed area) in Fig. \ref{F123}(c) closing at $w/\gamma=2/3$ [Eq. (\ref{eigs_b2}) with $s_{12}=0$]. We remark that the presence of different frequency regions and the related branching of frequencies are associated to the presence of EPs. For the sake of comparison in appendix \ref{appA} in Fig. \ref{eigs_fig} we show  the smooth eigenvalues variation with parameters in the absence of coalescence phenomena.

\begin{figure}[t!]
 \centering
 \includegraphics[width=0.9\columnwidth]{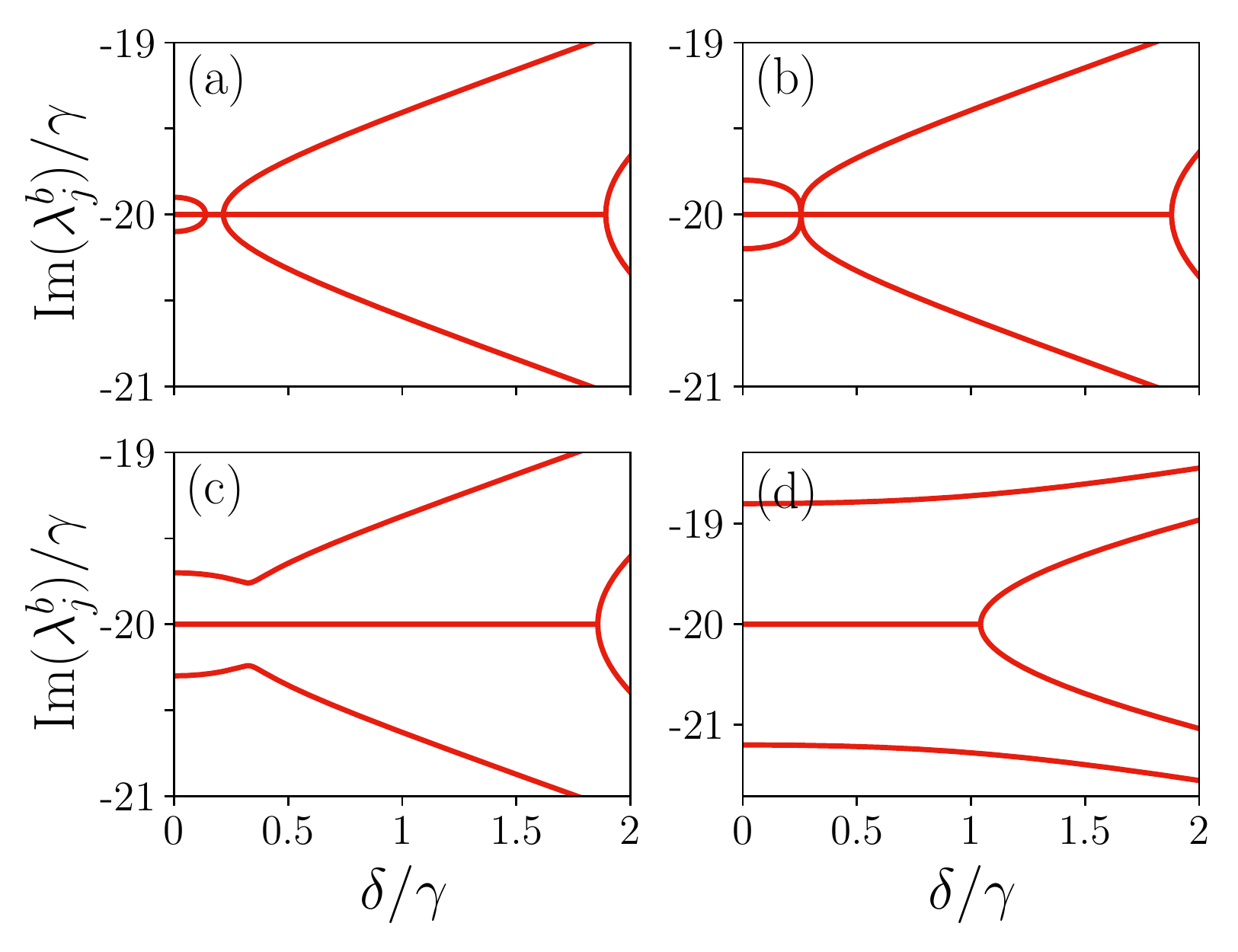}
 \caption{Eigenfrequencies for $w/\gamma=1$, $\omega_0/\gamma=20$, varying the detuning and for multiple coupling strengths. (a) $s_{12}/\gamma=0.1$, (b) $s_{12}/\gamma=0.2$, (c) $s_{12}/\gamma=0.3$ and (d) $s_{12}/\gamma=1.2$.}
 \label{EP_s12}
\end{figure}

EPs separate dynamical regimes characterized by a different number of frequencies and the richest scenario is found for $s_{12}/\gamma=0$ and varying $w/\gamma$ and $\delta/\gamma$ [Fig. \ref{F123}(g)] where three different regimes are found: SFR, TFR and FFR,
all of them separated by lines of second order EPs (white lines). On the other hand, numerical analysis  reveals that when $s_{12}/\gamma\neq0$ the system generally displays  four frequencies and four decay rates as EPs are not present
(as in Fig. \ref{eigs_fig}). A notable exception is the case of $w/\gamma=1$ in which up to three EPs can be found for $s_{12}/\gamma\geq0$ and $\delta/\gamma<2$. We start at $s_{12}/\gamma=0$ in which there are the two EPs that belong to the white lines of Fig. \ref{F123} (g), and an isolated EP at $\delta/\gamma=0$ [see Eq. (\ref{eigs_b2})]. As we increase $s_{12}/\gamma$ the two small detuning EPs approach each other until they annihilate at $\delta/\gamma\approx0.26$, $s_{12}/\gamma\approx0.21$, then only the large detuning EP remains. This last EP drifts to smaller $\delta/\gamma$ as the coupling is increased until it reaches $\delta/\gamma=0$ at $s_{12}/\gamma=\sqrt{2}$ [Eq. (\ref{eigs_b2})] and disappears for larger coupling strengths. This peculiar behavior is illustrated in Fig. \ref{EP_s12}.

Our results show that, considering the Hamiltonian (\ref{Ham}), coalescence is in general found in the absence of direct coupling between the spins. 
This raises the question of whether spins direct coupling always hinders coalescence or it depends on the nature of the interaction term. Consider for instance the case where the coupling has the form $s_z\hat{\sigma}_1^z\hat{\sigma}_2^z$, i.e. pure dephasing . 
This situation was analyzed for instance in Ref. \cite{PRB}, where the emission spectrum of two coupled quantum dots was studied.
In this case, one could still find  a double EP  at $w/\gamma=s_{12}/\gamma=0$ and $\delta/\gamma=1$, with the difference that now there are always at least two different frequencies, $\omega_0\pm 2s_z$. Another EP is found in the presence of pumping, when $\delta/\gamma=s_{12}/\gamma=0$, $w/\gamma=1$ and $s_z/\gamma=1/\sqrt{2}$. Then, the presence of pure dephasing would have strong effects on coalescence, changing completely the scenario displayed in Fig. \ref{F123}(g).

\section{Synchronization of the coherences}\label{sync}

In this section we analyze the synchronization in the dynamics of observables related to the spin coherences (living in the $\mathcal{L}_{b(c)}$ sectors). Synchronization emerges here as a transient monochromatic oscillation in which the coherences of both qubits remain phase-locked until they reach the non-oscillatory stationary state of the system. In fact, as anticipated, in this system we find that synchronous dynamics can appear due to two different mechanisms. The first we have reported above is coalescence, which occurs widely when $s_{12}/\gamma=0$ and enables the system to display just one frequency (SFR). As we show in Sec. \ref{sync_coal}, despite the fact that the coherences oscillate monochromatically from the beginning independently on the initial condition, phase-locking emerges generally after a transient time related to the decay rates of the eigenmodes of $\mathcal{L}_{b(c)}$. The second  mechanism is non-degenerate subradiance, and is also known in the literature as transient synchronization \cite{SyncRev2}.
This mechanism arises in our system provided that  $s_{12}/\gamma\neq0$ (Sec. \ref{sync_trad}). In this case, the coherences display in the early stage of the dynamics  four different frequencies.  However, the presence of a slowly decaying subradiant eigenmode leads to frequency selection and  brings the system to a regime where both frequency- and phase-locking are present after a transient time in which the rest of the eigenmodes decay out. 

\begin{figure*}[t]
 \centering
 \includegraphics[width=2\columnwidth]{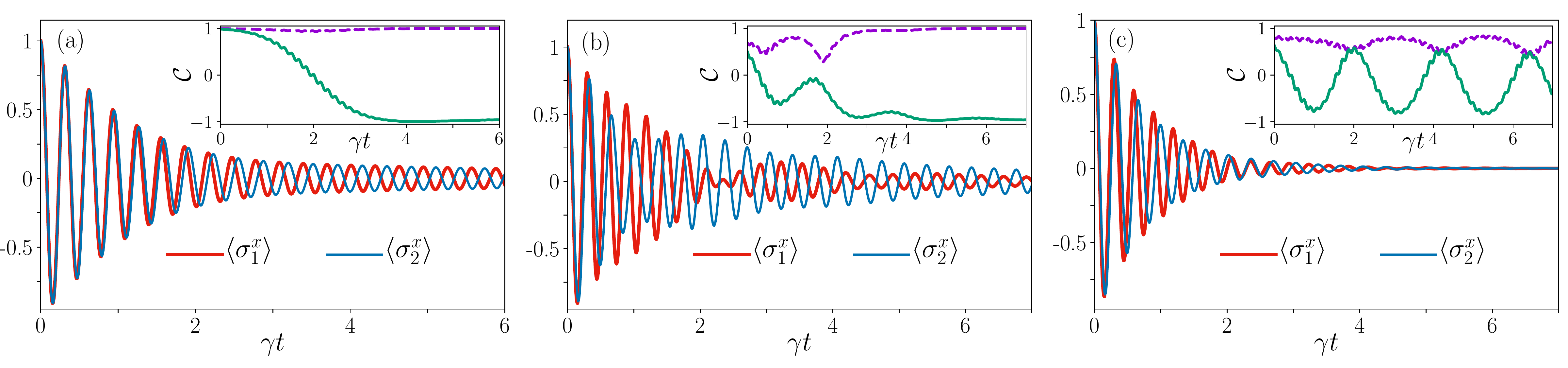}
 \caption{Main panels: $\langle\hat{\sigma}_1^x\rangle$ (red thick line) and $\langle\hat{\sigma}_2^x\rangle$ (blue thin line) for the initial condition $|\phi_0\rangle=(|ee\rangle+|eg\rangle+|ge\rangle+|gg\rangle)/2$.  Insets: $\mathcal{C}_{\langle \hat{\sigma}_1^x(\gamma t)\rangle,\langle \hat{\sigma}_2^x(\gamma t)\rangle}(\gamma \Delta t)$ (green solid line) and $\mathcal{C}_{\text{max}}$ (purple dashed line) with $\Delta t=1.2/\gamma$ and  delay range $\delta\tau=0.35/\gamma$. (a) SFR with parameters $\omega_0/\gamma=20$, $s_{12}/\gamma=0$, $w/\gamma=0.1$, $\delta/\gamma=0.3$. (b) FFR with parameters $\omega_0/\gamma=20$, $s_{12}/\gamma=1$, $w/\gamma=0.1$, $\delta/\gamma=2$. (c) Same as in (b) but fixing the incoherent pumping rate to $w/\gamma=0.75$.} 
 \label{Ftraj}
\end{figure*}

\subsection{Synchronization due to coalescence}\label{sync_coal}

To start with, let us consider 
the phenomenon of synchronization due to coalescence, emerging in the SFR regime in which $\mathcal{L}_b$ has just one eigenfrequency and four decay rates. We will analyze the dynamics of $\langle \hat{\sigma}_{1,2}^x\rangle$, which display an oscillatory decay towards the stationary state, and  assess the emergence of synchronization with the use of the measures of synchronization introduced in appendix \ref{appB}, which are the Pearson factor (\ref{SyncMeasure}) and its maximized version,  optimized over all possible phase shifts. As we have anticipated, in spite of the presence of just one frequency, phase-locking between the coherences dynamics  is not guaranteed. This is evident in Fig. \ref{Ftraj}(a) in which the phase between the trajectories slips from zero to almost $\pi$ at $\gamma t\approx4$, where it remains locked until the oscillation completely decays out. The  Pearson factor accounting for delay (purple dashed line) is a good measure of the final synchronous oscillation, while we can appreciate the transient phase slip as signaled by the bare indicator (green solid line).

The slip of the relative phase can be understood by analyzing the semi-analytical solution of $\langle \hat{\sigma}_{1,2}^x\rangle$ (see appendix \ref{appA}). Indeed, we can particularize Eq. (\ref{solution}) to the SFR in which $\text{Im}(\lambda^b_k)=-\omega_0$ $\forall k$ and hence
\begin{equation}\label{sol_SFR}
\langle \hat{\sigma}_j^x(t)\rangle=\sum_{k=1}^4 2|p^b_{0k}\langle\tau_k^b\rangle_{xj}|e^{\text{Re}(\lambda_k^b)t}\cos[\psi_{k,xj}^b-\omega_0t], 
\end{equation}
the coefficients being defined in the appendix and $j=1,2$. Importantly, both the weight ($p^b_{0k}$) and phase ($\psi_{k,xj}^b$) associated to each eigenvalue depend on the initial condition. Then from Eq. (\ref{sol_SFR}) we find that there are multiple terms oscillating at the same frequency but with a different phase. The relative importance of each term changes in time due to the time dependent part of the weight factor  
$e^{\text{Re}(\lambda_k^b)t}$, where the eigenvalues of $\mathcal{L}_b$  are ordered such that $\lambda_4^b$ is the one with the smallest real part  in absolute value. This makes the relative phase between the qubits to slip from the initial value determined by the initial condition to $\Delta \psi=\psi_{4,x1}-\psi_{4,x2}$ in a time scale related to $\text{Re}(\lambda_3^b)$, in which  all terms in Eq. (\ref{sol_SFR}) except the less damped one are no longer significant. Notice that, the more similar $\text{Re}(\lambda_{3,4}^b)$ are, the more damped will be the oscillations when the relative phase eventually locks. The dependence of the weights on the initial condition can be illustrated considering the same parameters as in Fig. \ref{Ftraj}(a) but  with the initial condition $|\phi_0\rangle=(|ee\rangle-|eg\rangle+|ge\rangle-|gg\rangle)/2$, in which is found that the relative phase is almost $\pi$ from the beginning (not shown here).

It is also interesting to comment on the general effect of increasing  the incoherent pumping rate $w/\gamma$. As we have shown in Fig. \ref{F123}(g) the SFR involves a wide range of values of $w/\gamma$, which implies that the same synchronization mechanism is present for large $w/\gamma$. Nevertheless, notice that the decoherence rate increases significantly with $w/\gamma$ (as also appreciated in panel (e) of the same figure), damping strongly the coherent oscillations of $\langle \hat{\sigma}_{1,2}^x\rangle$. Thus, the amplitude of the synchronous oscillation decreases significantly with increasing incoherent pumping, which makes the phenomenon  harder to be observed and finally hinders it.

\subsection{Synchronization due to non-degenerate subradiance}\label{sync_trad}

Here we analyze the case without degeneracy where multiple frequencies are present ($s_{12}/\gamma\neq0$) since the early stage of the dynamics. In this parameter regime, spontaneous synchronization can emerge leading to a monochromatic evolution and it is known to be related to the presence of a non-degenerate subradiant eigenmode \cite{Bellomo}.
In this case $\mathcal{L}_b$ generally displays  four frequencies and four different decay rates, and thus synchronization can only emerge in the presence of a slowly dissipating eigenmode \cite{SyncRev2}, i.e. when $\text{Re}(\lambda^b_4)/\text{Re}(\lambda^b_3)\ll1$, which leads to frequency selection. This statement can be understood analyzing the semi-analytical solution of $\langle \hat{\sigma}_{1,2}^x\rangle$ given in Eq. (\ref{solution}): at the beginning the four different frequencies are involved and thus the qubits oscillate irregularly, however, as each frequency component decays with a different rate given by $\text{Re}(\lambda^b_k)$, after a transient time, if $\text{Re}(\lambda^b_4)/\text{Re}(\lambda^b_3)\ll 1$, there is a significant oscillation governed by the eigenmode with the smallest decay rate $\text{Re}(\lambda^b_4)$, making the qubits to oscillate synchronously with the phase difference locked to $\Delta \psi=\psi_{4,x1}-\psi_{4,x2}$. An example of such phenomenon is shown in Fig. \ref{Ftraj}(b), where we can observe that after a time of about $\gamma t\approx 4$ the two qubits oscillate synchronously with a difference of phase of about $\pi$. Both indicators of synchronization, the Pearson factor and the maximized one, reach a stationary value close to -1 or 1 respectively. When comparing panels (a) and (b), we notice that in both cases synchronization emerges after a transient of a similar duration and the lasting amplitudes are of similar magnitude. Nevertheless, in the latter case the transient to synchronization displays strong amplitude modulations related to the presence of multiple frequencies. 

The influence of the different parameters on the synchronization behavior can be analyzed systematically by studying the ratio of the two smallest eigenvector decay rates  \cite{SyncRev2}. Indeed, the case with $w/\gamma=0$ was already studied in Ref. \cite{Bellomo}, in which it was shown that the more detuned are the qubits, the more coherent coupling is needed  for synchronization to emerge, analogously to the classical Arnold-tongue behavior. As a matter of fact, we find that a nonzero $w/\gamma$ preserves this overall behavior but decreases the capacity of the qubits to synchronize. This is illustrated in Fig. \ref{Ftraj}(c), where the increased  incoherent pumping rate prevents the emergence of synchronization, as indicated by  the marked oscillatory behavior of the Pearson factor. 

\begin{figure}[t!]
 \centering
 \includegraphics[width=0.9\columnwidth]{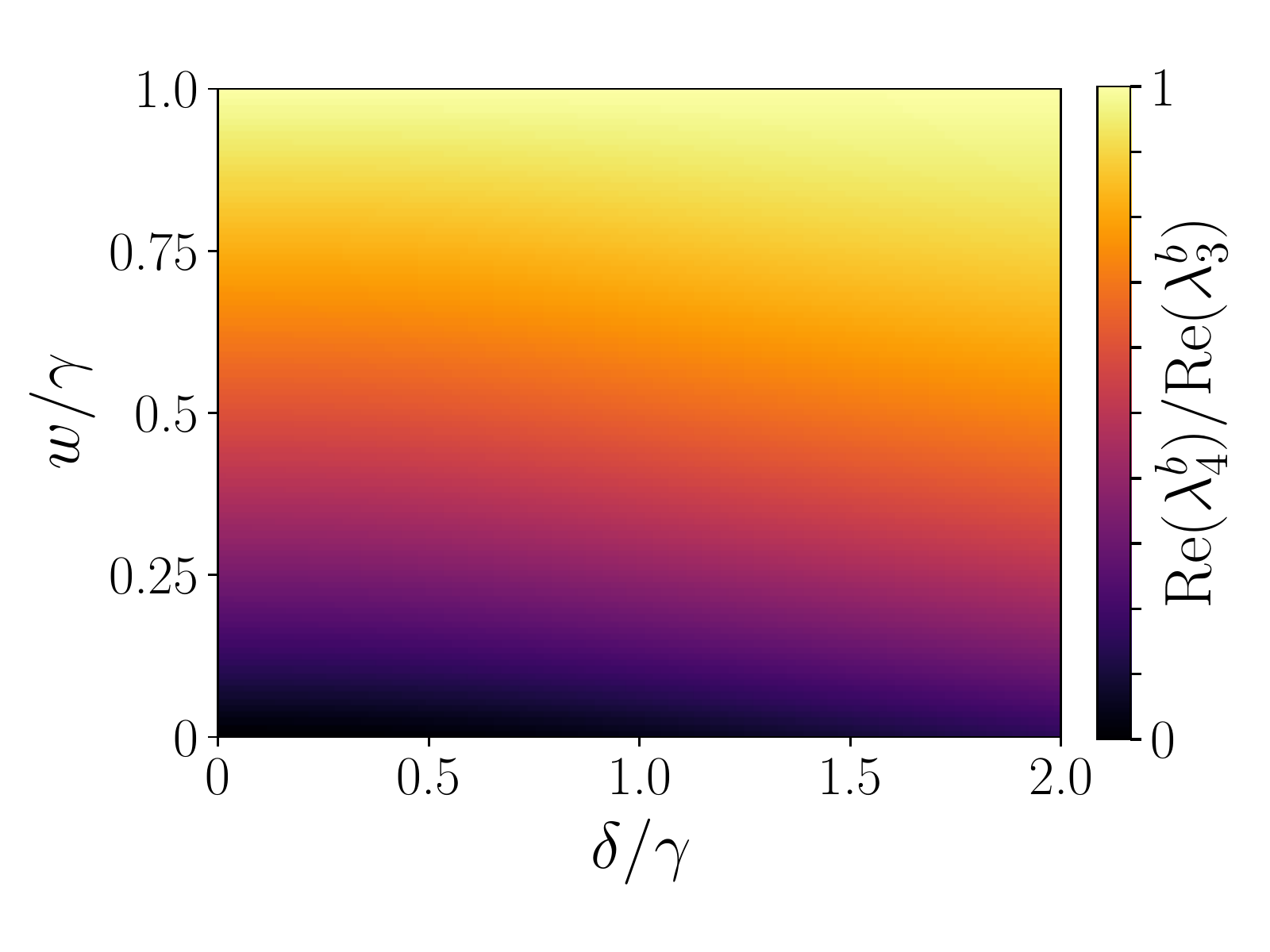}
 \caption{In color: ratio of the two smallest decay rates $\text{Re}(\lambda^b_4)/\text{Re}(\lambda^b_3)$ varying $w/\gamma$ and $\delta/\gamma$, with the other parameters fixed to $\omega_0/\gamma=20$ and $s_{12}/\gamma=1$.}
 \label{map_ratio}
\end{figure}

The detrimental effect of the incoherent pumping can be understood by recalling that it constitutes an additional decoherence channel acting locally on each qubit, and thus as $w/\gamma$ is increased, the effect of the common environment is counteracted by local decoherence which decreases the disparity between the two smallest decay rates. This is explicitly shown in Fig. \ref{map_ratio} in which the ratio of the two smallest eigenvector decay rates is plotted varying $w/\gamma$ and $\delta/\gamma$. For small enough $w/\gamma$ we can see that there is one decay rate significantly smaller than the rest enabling the emergence of synchronization [as in Fig. \ref{Ftraj}(b)]. However, as  $w/\gamma$ increases this ratio tends to one and synchronization no longer emerges [as in Fig. \ref{Ftraj}(c)]. Moreover, notice that the overall magnitudes of the decay rates increase with $w/\gamma$ causing also a faster damping of the coherent oscillations, as we have also commented in Sec. \ref{sync_coal}. 

Notice that in our system, the two  kinds of synchronization cannot emerge in the same parameter regime. This is so, as when $s_{12}/\gamma=0$
and EPs are predicted, $\mathcal{L}_b$ either displays a single frequency (SFR) or displays several frequencies with the same decay rate (TFR and FFR). Moreover, it turns out that in the TFR the smallest decay rate is the one shared by two frequencies making not possible the emergence of synchronization by the second mechanism.

We remark that actually also coalescence (beyond the singular points) displays a larger damping in one mode than in another, but both share the same oscillation frequency. In other words,  coalescence is accompanied by sub/superradiance 
in the presence of \textit{frequency degeneracy} and enables the  emergence of phase-locking, reported in the previous section. This is of course different from having two oscillating modes at different frequencies and for this reason we refer here specifically to \textit{non-degenerate} subradiance. More details on this last point are presented in the next section in which the correlation spectrum is considered.

\section{Signatures of synchronization in the correlation spectrum}\label{spectrum}

In this section we present a complementary view of the phenomenon of synchronization, analyzing its signatures in the two-time correlation spectrum, an indicator relevant when probing the system and accessible in many setups. This approach to characterize synchronization was taken  for instance in Ref.  \cite{Holland1}, and as we will show it serves to illustrate the relation between synchronization and super/subradiance phenomena. The  correlations considered here lie in the same Liouvillian sectors $\mathcal{L}_{b(c)}$ as the local observables considered in the previous section. Two-time correlations can be considered either for collective  spin operators $\langle \hat{L}(t+\tau)\hat{L}^\dagger(t)\rangle$ or for local  ones, $\langle\hat{\sigma}_{j}^-(t+\tau)\hat{\sigma}^+_{j}(t)\rangle$. An important motivation behind considering both  collective and local correlations  comes from the master equation in Eq. (\ref{ME}), in which  both kind of operators are present in the dissipators  $\mathcal{D}$, in form of collective dissipation or local pumping. Let us proceed as follows: first we will consider the case $w/\gamma=0$ in Sec. \ref{specw0}, where analytical results can be obtained and can be used to illustrate our main results, then, the role of incoherent pumping will be discussed \ref{specw}. The mathematical details are presented in appendix \ref{appC}. 

\subsection{Case with $w/\gamma=0$}\label{specw0}

We consider the system in the  absence of pumping  ($w/\gamma=0$) for both kinds of synchronization regimes discussed in the previous section. We consider both $\langle \hat{L}(\tau)\hat{L}^\dagger(0)\rangle_{ss}$, and $\langle\hat{\sigma}_{j}^-(\tau)\hat{\sigma}^+_{j}(0)\rangle_{ss}$, where the subscript $ss$ indicates they are computed in the stationary state of the system, which in the absence of driving is $|gg\rangle\langle gg|$. This is the reason why the calculation can be done analytically, just considering the one excitation sector of $\mathcal{L}_b$ as shown in  appendix \ref{appC}. The  Fourier transform of these two-time correlations  [Eq. (\ref{out_spec})], or correlation spectrum, displays the relevant information about the collective excitations of the system, such as their frequency, decay rate and overlap of the correlators with the eigenmodes.

\begin{figure}[t!]
 \centering
 \includegraphics[width=0.95\columnwidth]{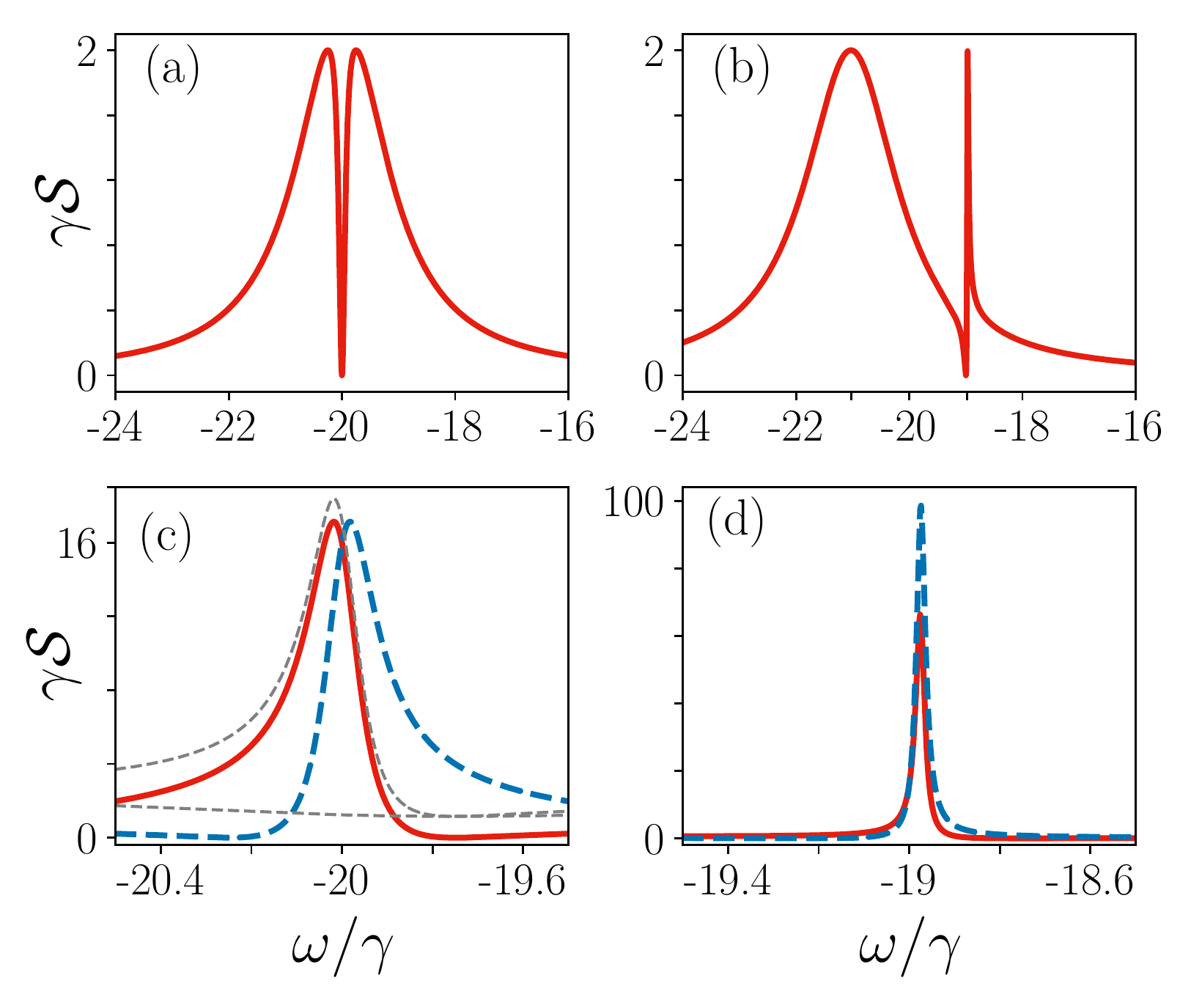}
 \caption{Fourier transform of $\langle \hat{L}(\tau)\hat{L}^\dagger(0)\rangle_{ss}$ (a) and (b) and of  $\langle\hat{\sigma}_{1(2)}^-(\tau)\hat{\sigma}^+_{1(2)}(0)\rangle_{ss}$ (c) and (d) in red solid (blue dashed) lines. The parameters are fixed to $\omega_0/\gamma=20$, $\delta/\gamma=0.5$  with $w/\gamma=0.0$ in all figures. In (a) and (c) we have $s_{12}/\gamma=0$, in (b) and (d) $s_{12}/\gamma=1$. In the case $s_{12}/\gamma=1$, the broad resonance has frequency $-\omega_0-V_I/2=-21.025\gamma$ and width $(\gamma+V_R)/2=0.988\gamma$, while the narrow one $-\omega_0+V_I/2=-18.975\gamma$ and $(\gamma-V_R)/2=0.024\gamma$. Notice that these frequencies are indicated in panel (b) as the ticks without label. In panel (c) we have included in gray dashed lines the two terms of Eq. (\ref{spec_S1}) that when subtracted yield the red curve.}
 \label{specsw0}
\end{figure}

We start considering the correlation spectrum for collective operators $\mathcal{S}_{\hat{L}\hat{L}^\dagger}(\omega)$ in the SFR [Fig. \ref{specsw0}(a)]
induced by coalescence  and in a case in which synchronization emerges in the presence of   non-degenerate subradiance [Fig. \ref{specsw0}(b)]. In both cases we observe signatures of super- and subradiant behavior, the latter being related to the eigenmode synchronizing the qubits either in the presence of coalescence or when there are multiple frequencies. Moreover,
interference effects are also present as the spectrum is not simply Lorentzian. However, in the SFR the interference occurs just at the resonance frequency $\omega_0/\gamma$, while for non-degenerate subradiance  the interference occurs between two resonances of different frequency that correspond to $\omega_0\pm \text{Im}(V)/2$. Notice that in all these plots, when comparing the two regimes, the frequency window of the plots is taken of the same magnitude such that the width of the peaks can be compared faithfully.

Considering the exact expressions for $\mathcal{S}_{\hat{L}\hat{L}^\dagger}(\omega)$ we find that, in the SFR ($s_{12}/\gamma=0$) 
\small
\begin{equation}\label{spec_C1}
\begin{split}
\mathcal{S}_{\hat{L}\hat{L}^\dagger}(\omega)=\frac{2}{V}\bigg[\frac{(\omega+\omega_0)^2}{(\omega+\omega_0)^2+\frac{1}{4}(\gamma-V)^2}\\
-\frac{(\omega+\omega_0)^2}{(\omega+\omega_0)^2+\frac{1}{4}(\gamma+V)^2}\bigg].
\end{split}
\end{equation}
\normalsize
This corresponds to two superposed (interfering) resonances, opposite in sign and  each centered at the same frequency $\omega_0$ but with a different decay rate (in this case $V$ is real), which yield a broad peak with a transparency window whose width is given by the narrow resonance. Notice that the width of the narrow and broad resonance can be quite disparate for small enough detuning, leading to pronounced \textit{degenerate} subradiant and superradiant eigenmodes (as found also in Ref. \cite{PRB}), the former enabling phase-locking of the coherences. Moreover, the multiplying factor $(\omega+\omega_0)$ implies that $\mathcal{S}_{\hat{L}\hat{L}^\dagger}(-\omega_0)=0$, as observed in the plots. 

In the case $s_{12}/\gamma\neq0$, $V$ becomes complex, and we denote its real and imaginary parts as $V_R$ and $V_I$ respectively. The exact results now read as
\small
\begin{equation}\label{spec_C2}
\begin{split}
\mathcal{S}_{\hat{L}\hat{L}^\dagger}(\omega)=\frac{2(\omega+\omega_0-s_{12})}{|V|^2}\bigg[\frac{\gamma\frac{V_I}{2}+V_R(\omega+\omega_0-V_I)}{(\omega+\omega_0-\frac{V_I}{2})^2+\frac{1}{4}(\gamma-V_R)^2}\\
-\frac{\gamma\frac{V_I}{2}+V_R(\omega+\omega_0+V_I)}{(\omega+\omega_0+\frac{V_I}{2})^2+\frac{1}{4}(\gamma+V_R)^2}\bigg],
\end{split}
\end{equation}
\normalsize
in which we observe again the interference of two resonances, but now centered at different frequencies $\omega=\omega_0\pm V_I/2$ and with different decay rates.  Notice that here completely destructive interference occurs at $\omega=-\omega_0+s_{12}$. Moreover,  for $s_{12}/\delta\gg1$, $V_R\approx\gamma$ while $V_I\approx2 s_{12}$, which implies that there is a significantly superradiant eigenmode and a significantly subradiant one, the latter being the one synchronizing the spins. This is clearly observed in Fig. \ref{specsw0} (b), in which the superradiant eigenmode is centered around $\omega\approx-\omega_0-s_{12}$ and the subradiant one at around $\omega\approx-\omega_0+s_{12}$.

We now compare these results with the ones for local correlation spectra (for each spin)  $\langle\hat{\sigma}_{1(2)}^-(\tau)\hat{\sigma}^+_{1(2)}(0)\rangle_{ss}$ for the same two cases [see Fig. \ref{specsw0} (c),(d)]. Focusing first in the SFR, we observe that $\mathcal{S}_{\hat{\sigma}_{1(2)}^-\hat{\sigma}_{1(2)}^+}(\omega)$ displays an asymmetric peak slightly displaced at the left (right) of $\omega_0$. This is still an interference effect as  the exact results show:
\small
\begin{equation}\label{spec_S1}
\begin{split}
\mathcal{S}_{\hat{\sigma}_{1(2)}^-\hat{\sigma}_{1(2)}^+}(\omega)=\frac{2}{V}\bigg[\frac{(\omega+\omega_0)[\omega+\omega_0\mp\frac{\delta}{2}]+\frac{\gamma}{4}(\gamma-V)}{(\omega+\omega_0)^2+\frac{1}{4}(\gamma-V)^2}\\
-\frac{(\omega+\omega_0)[\omega+\omega_0\mp\frac{\delta}{2}]+\frac{\gamma}{4}(\gamma+V)}{(\omega+\omega_0)^2+\frac{1}{4}(\gamma+V)^2}\bigg],
\end{split}
\end{equation}
\normalsize
 where the upper sign corresponds to spin 1 and the lower sign to spin 2. 
 Here we find  the peaks of each spin to be centered at slightly shifted frequencies: the two time correlations of each spin are affected by the presence of the other one, that is detuned, and each spectrum experiences a pushing effect. Of course these self-correlations enter also in the collective spectra described above but there the cross-correlations between spins also play a major role. In this case we have plotted each term of Eq. (\ref{spec_S1}) in gray dashed lines in Fig. \ref{specsw0}(c) from which we can appreciate that the term with the small decay rate already accounts for the very asymmetric resonance,  while the contribution from the other term is almost homogeneous.

 In the case of $s_{12}/\gamma\neq0$, Fig. \ref{specsw0}(d),  we see that the self-correlations mainly display the sharp peak also present in the collective spectrum of correlations: the superradiant eigenmode is barely visible   in this case while the subradiant one -- which leads to synchronization -- is the main contribution. The main reason for the difference between Figs. \ref{specsw0}(d) and  \ref{specsw0}(b) is that the collective operator in the former is almost orthogonal to the subradiant eigenmode. In fact $\hat{L}$ is exactly the superradiant eigenmode in the absence of detuning. Therefore, the contributions of both eigenmodes acquire the same importance in the collective spectrum. In this case the analytical results are too cumbersome to provide additional insights.

One of the main results discussed here is that the two kinds of synchronization present different signatures in the correlation spectrum. In the case of synchronization due to coalescence, we find an interference effect at the resonance frequency, which manifests itself as a visible dip in the case of collective measurement or as asymmetric resonances when addressing each spin separately. On the other hand, in the  presence of non-degenerate subradiance, synchronization is signaled by an asymmetric spectrum with significant disparity in the width of the resonances, which is visible both in collective measurements (when $\hat{L}$ is not orthogonal to the subradiant eigenmode) and in local correlations, where he subradiant mode turns out to dominate.

\begin{figure}[t!]
 \centering
 \includegraphics[width=1\columnwidth]{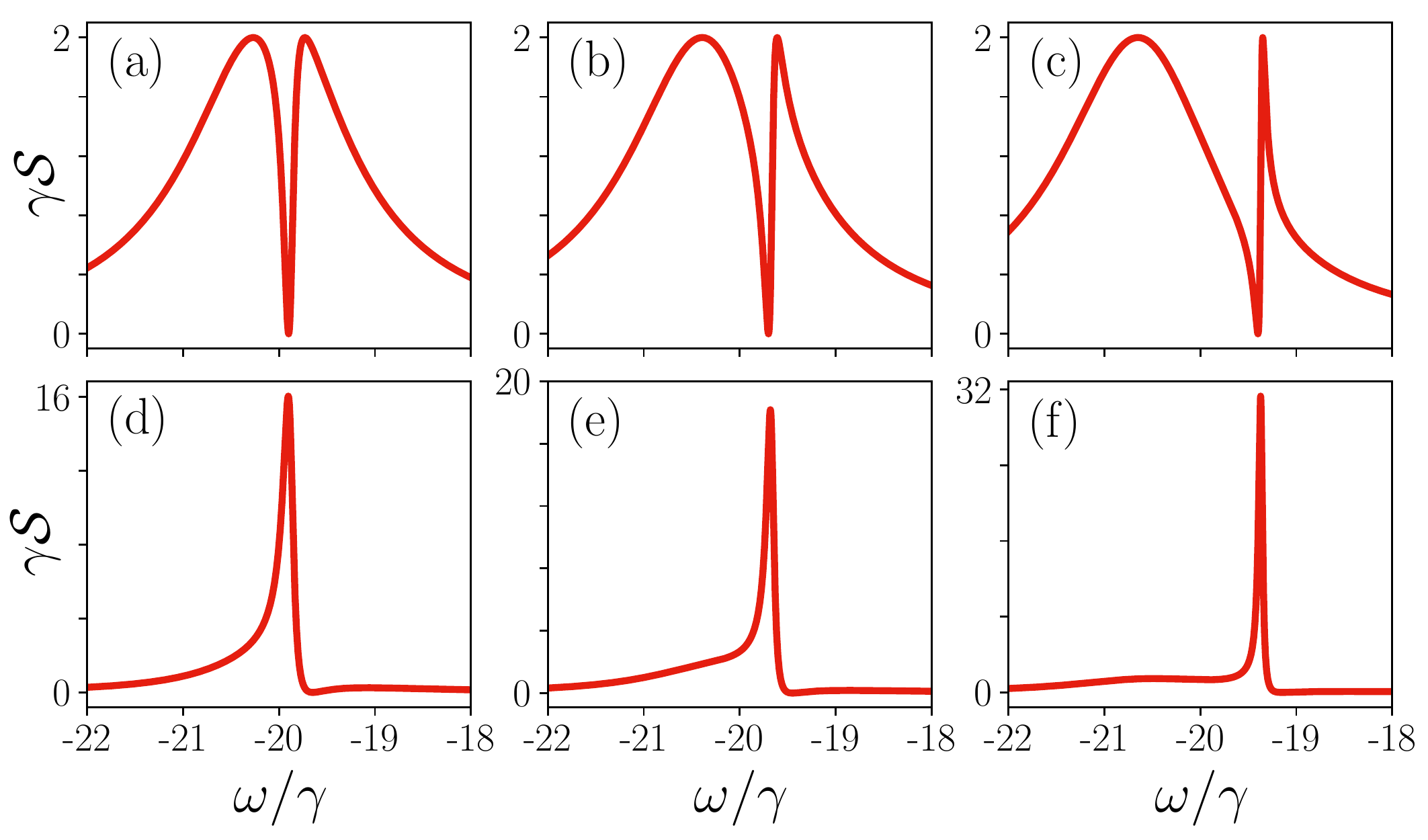}
 \caption{Fourier transform of $\langle \hat{L}(\tau)\hat{L}^\dagger(0)\rangle_{ss}$ (a), (b), (c), and of  $\langle\hat{\sigma}_{1}^-(\tau)\hat{\sigma}^+_{1}(0)\rangle_{ss}$ (d), (e), (f). The parameters are fixed to $\omega_0/\gamma=20$, $\delta/\gamma=0.5$  with $w/\gamma=0$. In (a), (d) $s_{12}/\gamma=0.1$, in (b), (e) $s_{12}/\gamma=0.3$, and in (c), (f) $s_{12}/\gamma=0.6$.  }
 \label{specsw0_barr}
\end{figure}

The evolution of the spectrum as the coupling is progressively increased can be appreciated  in Fig. \ref{specsw0_barr}, complementing Fig. \ref{specsw0}. In panels (a) to (c) we observe how, as the coherent coupling strength increases, the symmetric interference window deforms progressively yielding two resonances of different frequency and width. Notice also how the node in the spectrum departs progressively from $-\omega_0$. This is also observed for self-correlations, (d) to (f), in which from the initial asymmetric resonance a broad (hard to appreciate)  resonance and a narrow one    (behind the appearance of synchronization) emerge. Looking at the transition from \ref{specsw0}(a) to \ref{specsw0_barr}(a) to (c), we observe that, while the presence of exceptional points represents a singular scenario and is immediately lost if the parameters are modified, the physical effects of such changes are continuous. Indeed, the spectrum of Eq. (\ref{spec_C2}) tends to the one of (\ref{spec_C1}) in the limit of infinitesimal coupling.

An interesting point is that, in general, we find that the interference effects introduce a fine structure in the spectrum of the system, which displays features of width smaller than the intrinsic one given by $\gamma$: as transparency windows, subradiant eigenmodes, or completely destructive interferences. Indeed, interference effects in the spectrum of quantum systems can be exploited, for instance,  in laser cooling schemes as  described in \cite{Morigi}.

\subsection{Case with {$w/\gamma\neq0$}}
\label{specw}

\begin{figure}[t!]
 \centering
 \includegraphics[width=0.9\columnwidth]{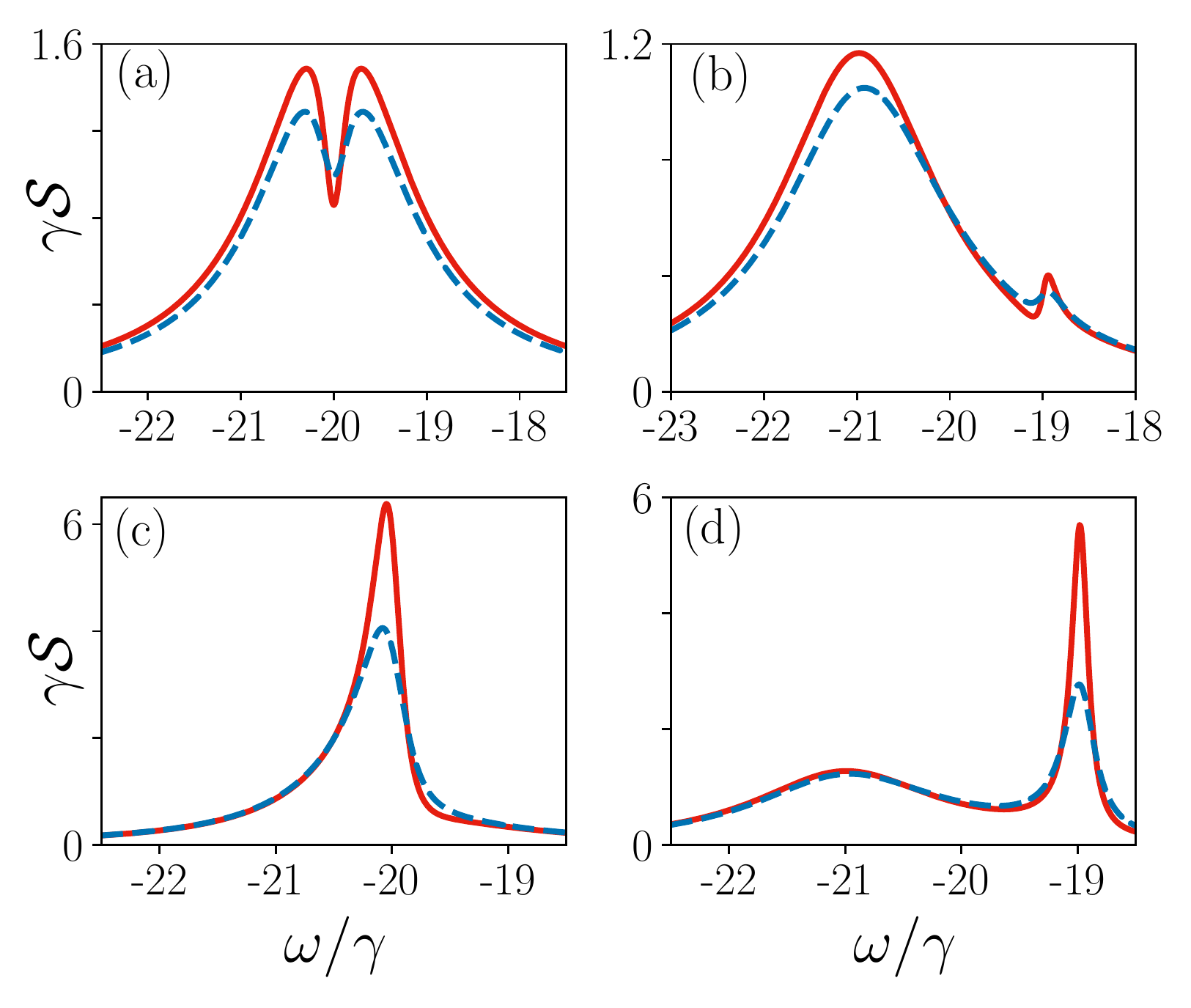}
 \caption{Fourier transform of $\langle \hat{L}(\tau)\hat{L}^\dagger(0)\rangle_{ss}$ (a) and (b) and of  $\langle\hat{\sigma}_{1}^-(\tau)\hat{\sigma}^+_{1}(0)\rangle_{ss}$ (c) and (d). The parameters are fixed to $\omega_0/\gamma=20$, $\delta/\gamma=0.5$ in all figures with $w/\gamma=0.05$ in red solid lines and $w/\gamma=0.1$ in blue dashed lines. In (a) and (c) we have $s_{12}/\gamma=0$, in (b) and (d) $s_{12}/\gamma=1$.}
 \label{specsw}
\end{figure}

In this section we address the effects of the incoherent pumping on the correlation spectrum. In this case the stationary state of the system is not the vacuum and involves in general all the density matrix elements of the sector $\mu=a$ \cite{Marco2,Bellomo}. The main results are illustrated in  Fig. \ref{specsw}, in which the spectrum of the collective and local correlations are plotted for two different values of $w/\gamma$. The results should be compared with those of the previous section, as we have just added a finite   incoherent pumping rate. In the SFR [panels (a) and (c)] we see that the main effect of the incoherent pumping is to decrease the visibility of the interference effects {and to reduce the disparity between super- and subradiant modes}; for the collective correlation the depth of the central dip decreases and its width increases,
while for local correlations the resonance becomes less asymmetric. In the case of $s_{12}/\gamma=1$ [panels (b) and (d)], we see that the width of the subradiant eigenmode increases significantly. Indeed, for the collective correlation the corresponding peak becomes barely visible, while for the local correlations it still dominates but with a significant decrease (increment) of the height (width) [compare with Fig. \ref{specsw0}(d)]. The increment of the width of the subradiant eigenmode  is already found and well illustrated in the expressions for the eigenvalues with $\delta/\gamma=0$, Eq. (\ref{eigs_b2}), in which we see that the real part of $\lambda_4^b$ increases linearly with $w/\gamma$, being completely subradiant for $w/\gamma=0$. This is a clear manifestation of the fact, commented above, that this local incoherent process counteracts collective dissipation, the latter being the mechanism behind strong disparities in the decay rates of the eigenmodes, which, as we have shown in Figs. \ref{specsw0} and \ref{specsw0_barr}, can lead to significant superradiant and subradiant effects in the correlation spectrum.

\section{Many-qubit scenarios}\label{manyqubit}

The aim of this section is to assess whether we can find synchronization due to coalescence and non-degenerate subradiance in more complex scenarios or in presence of different kinds of coherent and dissipative interactions. For this reason we will explain in detail some connections with the literature and introduce some possible many-qubits generalizations of the simple two-qubit system studied so far. In most cases, we will restrict the analysis of these many-qubit scenarios to the one-excitation sector, as it is enough to illustrate our point, and a more thorough study is out of the scope of this work. Some mathematical details are worked out in Appendix \ref{appd}.

As a starting point, we comment that synchronization
has been found in a system of two detuned atomic clouds externally pumped and interacting with a cavity 
mode in the bad-cavity limit, which acts as an effective collective dissipation channel
\cite{Holland1}. This system constitutes a possible many-qubit generalization of our 
two-qubit model, and synchronization is actually due to coalescence.  The master equation 
for the atomic clouds can be written as 
\begin{eqnarray}\label{ME2}
\dot{\hat{\rho}}&=&-i\sum_{j=1}^N\big[\frac{\omega_1}{2}\hat{\sigma}^z_{Aj}+\frac{\omega_2}{2}\hat{\sigma}^z_{Bj},\hat{\rho}\big]+\gamma_c\mathcal{D}[\hat{J}^-]\nonumber\\
&+&w\sum_{j=1}^N(\mathcal{D}[\hat{\sigma}_{Aj}^+]+\mathcal{D}[\hat{\sigma}_{Bj}^+]),
\end{eqnarray}
in which we can define the detuning between the two clouds of atoms, A and B, as $\delta=\omega_1-\omega_2$, and the collective dissipation channel with $\hat{J}^-=\sum_{j=1}^N(\hat{\sigma}_{Aj}^-+\hat{\sigma}_{Bj}^-)$ and collective decay rate $\gamma_c$, which depends on the cavity decay rate and atom-cavity interaction strength \cite{Holland1}.
In particular, it is shown that this system can reach a regime in which collective emission of light is  just at one frequency, $\omega_0=(\omega_1+\omega_2)/2$, despite a nonzero detuning  between the two atomic clouds. This striking behavior is identified in Ref. \cite{Holland1} as a form of synchronization of the atomic system, in which the transition between the synchronized and the unsynchronized regimes is studied looking at the behavior of the  first-order correlation function between the two clouds. According to our analysis, we can state that this transition point turns out to be an EP of the non-Hermitian matrix  governing the emission of light of the system. Thus, we can establish a connection between the many-body physics of \cite{Holland1} and the results discussed here.
\newline For vanishing driving $w=0$, the model (\ref{ME2}) is also suitable for an exact analytical treatment from which one can obtain the eigenvalues  governing the dynamics of the coherences $\langle\hat{\sigma}_{A(B)j}^-\rangle$ in the one-excitation sector, which are relevant when weakly probing the system:

\begin{eqnarray}
\lambda_{D,\pm}&=&-i(\omega_0\pm \frac{\delta}{2}),\\
\lambda_{B,\pm}&=&-i\omega_0-\frac{N\gamma_c}{2}\pm\frac{1}{2}\sqrt{(N\gamma_c)^2-\delta^2}, 
\end{eqnarray}

\noindent where the $\lambda_{D,\pm}$ display degeneracy with a multiplicity of $N-1$ each. Here, the indexes $D$ and $B$ stand, respectively for ``dark" and ``bright". Indeed, we see that there are two large completely dark bands of non-decaying modes, in which the eigenvalues are purely imaginary, and two bright modes with frequency and decay rate indicated by the imaginary and real part of $\lambda_{B,\pm}$, respectively. Interestingly, we find coalescence of the bright modes, with the EP located at $|\delta_c|=N\gamma_c$. Moreover, notice that for $|\delta|\ll|\delta_c|$ significant superradiant and subradiant effects are also present in these bright modes. However, in contrast to the previous scenarios, synchronization does not emerge for $w=0$. This is because only two modes $\lambda_{B,\pm}$ coalesce, while there are two large bands with different frequencies $\lambda_{D,\pm}$ that do not decay. This feature of $\lambda_{D,\pm}$ to be purely imaginary,  prevents dissipation to select one mode that synchronizes the system and thus the dynamics generally contains multiple frequencies at all times.

We now present two possible  scenarios of chains of qubits and different forms of dissipation and analyze the dynamics of the coherences in the one-excitation sector. As a first example,  we have $N$ unit cells each corresponding to our qubit model with $s_{12}/\gamma=0$ and $w/\gamma=0$. Furthermore, a dissipative term involving  nearest-neighbour qubits of different cells is present. This model is described by the following master equation:
\small
\begin{eqnarray}\label{ME3}
&\dot{\hat{\rho}}=-i\sum_{j=1}^N\big[\frac{\omega_1}{2}\hat{\sigma}^z_{Aj}+\frac{\omega_2}{2}\hat{\sigma}^z_{Bj},\hat{\rho}\big]+\gamma\sum_{j=1}^N\mathcal{D}[\hat{\sigma}^-_{Aj}+\hat{\sigma}^-_{Bj}]\nonumber\\
&+\gamma\sum_{j=1}^{N-1}\mathcal{D}[\hat{\sigma}^-_{Bj}+\hat{\sigma}^-_{A(j+1)}]+\gamma(\mathcal{D}[\hat{\sigma}^-_{A_1}]+\mathcal{D}[\hat{\sigma}^-_{B_N}]).
\end{eqnarray}
\normalsize
where we have considered open boundary conditions. Notice that for open boundary conditions we need to consider local dissipation terms for the qubits at the ends of the array, which ensure uniform dissipation for all qubits. As outlined in Appendix \ref{appd}, we can write down the eigenvalues ruling the dynamics of $\langle \hat{\sigma}^-_{A(B)j}\rangle$ in the one-excitation sector. Their expression is our main result for this model and reads:
\begin{equation}
\lambda_{k_l,\pm}=-i\omega_0-\gamma\pm \sqrt{\gamma^2\cos^2 \big(\frac{k_l}{2}\big)-\frac{\delta^2}{4}}.
\label{lambdak}
\end{equation}
with $k_l={\pi l}/{(N+1/2)}, \quad l=1,\dots,N.$ Thus we find that this system presents an EP for each $k_l$ at the critical detunings given by:
\begin{equation}
\delta_{c_l}=2\gamma|\cos (k_l/2)| , \quad l=1,\dots,N, 
\end{equation}
independently of the size of the system. Moreover, for $|\delta|<\text{min}[\delta_{c_l}]$ we find that for all $\lambda_{k_l,\pm}$ the imaginary part is the same, $\omega_0$, while there are $2N$ decay rates. This means that for sufficiently small detuning there is a region  with just one frequency, and thus synchronization arises due to coalescence. Here the size of the synchronized region diminishes with the system size, as for increasing $N$ we have $k_N\to \pi$, which implies $\delta_{c_N}\to 0$.

In the second example we illustrate how synchronization due to coalescence or non-degenerate subradiance can emerge in the presence of local dissipation
and coherent interactions, which represents the spin analog of the  chain of quantum harmonic oscillators discussed in Ref. \cite{Cabot_EPL}. The master equation for this system reads:
\small
\begin{eqnarray}\label{ME4}
\dot{\hat{\rho}}=&-i[\hat{H}_{AB},\hat{\rho}\big]+\sum_{j=1}^N \big(\gamma_A\mathcal{D}[\hat{\sigma}^-_{Aj}]+\gamma_B\mathcal{D}[\hat{\sigma}^-_{Bj}]\big),\\
\nonumber \\
\hat{H}_{AB}=&\sum_{j=1}^N\big[ \frac{\omega_0}{2}(\hat{\sigma}^z_{Aj}+\hat{\sigma}^z_{Bj})+s_{AB}(\hat{\sigma}^+_{Aj}\hat{\sigma}^-_{Bj}+h.c.)\big]\nonumber\\
&+\sum_{j=1}^{N-1} s_{AB}(\hat{\sigma}^+_{A(j+1)}\hat{\sigma}^-_{Bj}+h.c.),\nonumber
\end{eqnarray}
\normalsize
where $h.c.$ stands for Hermitian conjugate. Here, we have an array of $N$ unit cells  of identical qubits all with the same (nearest-neighbour) coupling $s_{AB}$  while losses are staggered.
The case in which the qubits of each cell were detuned was analyzed  in \cite{Cabot_PRL}, where results beyond the one-excitation sector were  obtained, and synchronization without coalescence was reported. For identical spins and considering the one-excitation sector dynamics for the coherences, the corresponding eigenvalues read as (see Appendix \ref{appd}):
\begin{equation}
\lambda_{k_l,\pm}=-i\omega_0-\frac{\gamma_+}{2}\pm \sqrt{\frac{\gamma_-^2}{4}-4s_{AB}^2\cos^2\big(\frac{k_l}{2}\big)},
\label{lambdak2}
\end{equation}
where $\gamma_\pm=(\gamma_A\pm\gamma_B)/2$, and the $k_l$'s are defined as after Eq. (\ref{lambdak}). We find an EP for each $l$, independently of system size, at a critical coupling given by:
\begin{equation}
s_{c_l}=|\gamma_-/[4\cos(k_l/2)]|,  \quad l=1,\dots,N.
\end{equation}
Notice that this  leads to synchronization due to coalescence for $s_{AB}<\text{min}[s_{c_l}]$. In this case, however, 
there is a minimum that is independent of the number of unit cells given by $|\gamma_-|/4$. Indeed, as the system size increases, we find $s_{c_1}/|\gamma_-|\to 1/4$. 

\begin{figure}[t!]
 \centering
 \includegraphics[width=0.9\columnwidth]{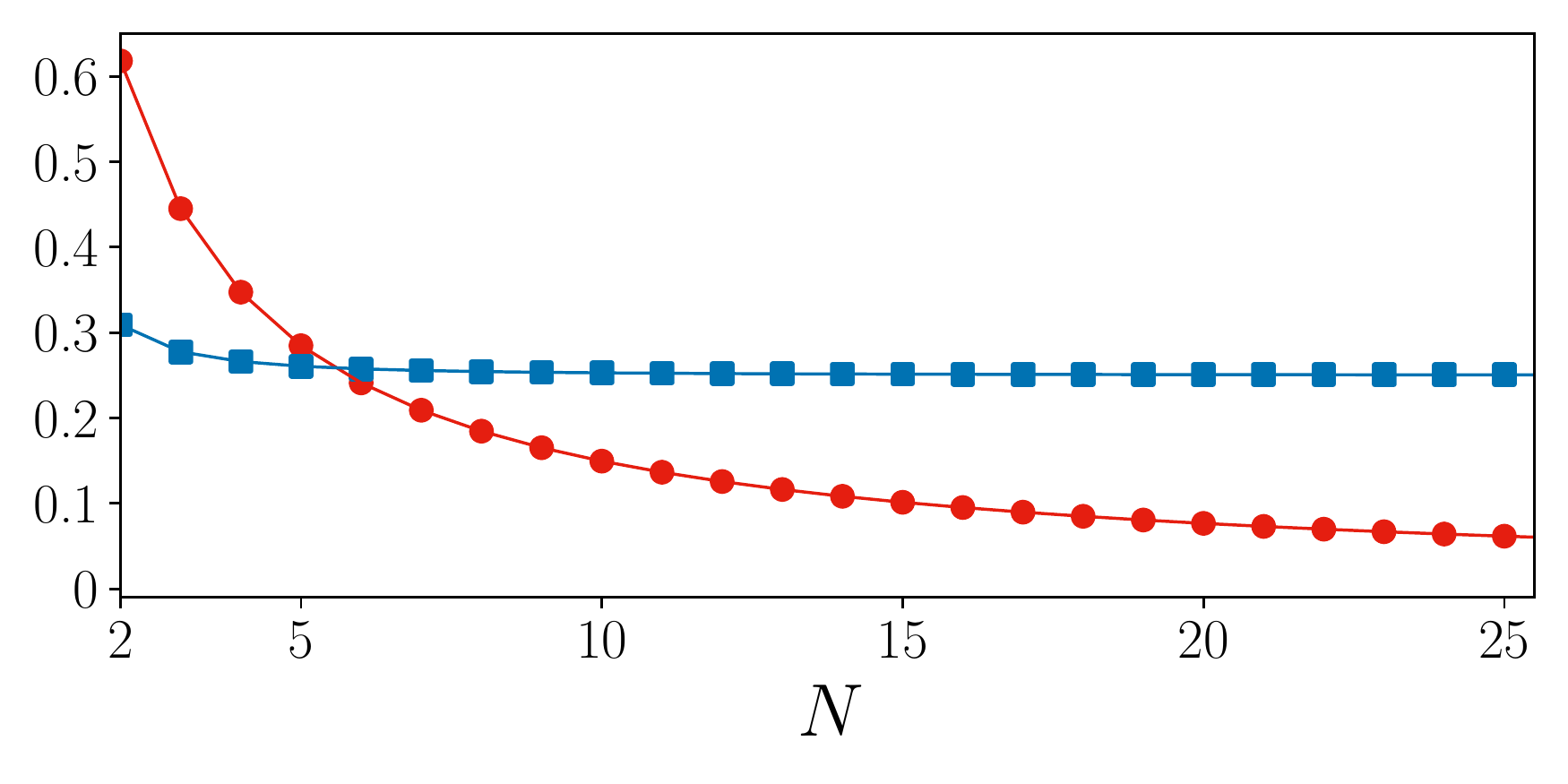}
 \caption{Regions of synchronization due to coalescence. Red circles: qubit chain with dissipative couplings. The line-points correspond to  $\delta_{c_N}/\gamma$ as the number of unit cells $N$ increases. For $|\delta|/\gamma$ below the line-points the system displays synchronization due to coalescence. Blue squares: qubit chain with coherent couplings. The line-points correspond to $s_{c_1}/|\gamma_-|$. In this case, for $s_{AB}/|\gamma_-|$ below the line-points the system displays synchronization due to coalescence.}
 \label{scaling}
\end{figure}

In Fig. \ref{scaling} we show how the size of the region with synchronization due to coalescence scales with the number of unit cells. In particular, for the array of dissipative couplings, this is indicated by $\delta_{c_N}/\gamma$ plotted in red circles, which diminishes with the size of the system. In the case of coherent couplings and local losses, this is indicated by $s_{c_1}/|\gamma_-|$ plotted in blue squares, which, interestingly, saturates to a minimum bound. This point was already noticed in \cite{Cabot_EPL} for a system of quantum harmonic oscillators, although the connection with synchronization was not established.

In summary, we have shown here  that the synchronization mechanisms explained in detail for the two-qubit model can emerge also in larger systems and even in the presence of different interactions. Thus, when these results are brought together, the physical picture is that the emergence of these phenomena does not depend on the very specific kind of interaction nor on the system size, but rather on combining appropriately a set of coherent and incoherent processes.

\section{Discussion and Conclusions}\label{discussion}

By analogy with what happens in classical systems,  quantum synchronization is connected to the spontaneous emergence of a monochromatic phase-locked oscillation among several coupled units. It is displayed by correlated local observables as well as in two-time correlation spectra \cite{Cabot_PRL}. 
In the framework of open quantum systems, this phenomenon  can be seen  as an ordered decay towards the stationary state of the system, and thus it is intimately related  to the presence of certain structure in the Liouvillian eigenspectrum of the system \cite{SyncRev2}.
This insight enables one  to establish a relation between synchronization and other phenomena such as subradiance \cite{Bellomo}, the presence of EPs, or to find signatures of this phenomenon in the correlation spectrum of the system as shown here or in \cite{Cabot_PRL}.

In this paper we have considered the  case of  detuned spins in the presence of collective dissipation and incoherent pumping and made a detailed comparison between two different mechanisms that can bring the system dynamics to a (quasi-) monochromatic behavior. 
One of these mechanisms is eigenvalue coalescence, in which collective excitations display the same frequency but multiple decay rates.
The second one is  non-degenerate subradiance, also known as transient synchronization,  in which, in the presence of multiple frequencies, one weakly damped long-lived collective excitation is responsible for the emergence of frequency selection and  phase-locking. While the emergence of synchronization due to non-degenerate subradiance was already known \cite{Bellomo},
we have established here the previously unnoticed 
 relationship between  the presence of exceptional points and quantum synchronization.

The two aforementioned mechanisms of synchronization are found to have different signatures in the correlation spectrum. As found in  chains of harmonic oscillators \cite{Cabot_EPL}, the signature of coalescence is an interference exactly at the resonant frequency, due to the presence of multiple eigenmodes with the same frequency but different decay rate. Here we have also found that, due to collective dissipation, this interference appears as a symmetric dip just at resonance, which in the abscence of coherent coupling, this is also a signature of super/subradiance \cite{PRB}. Indeed, as a general fact, coalescence can lead to super/subradiance, as at the
exceptional point two different damping rates emerge, their difference in magnitude being a signature of these phenomena. As the coherent coupling is turned on, the spectra of the SFR progressively split leading to two separate resonances with disparate widths, which are a clear indicator that  synchronization emerges now in the presence of multiple frequencies and due to the presence of a weakly damped non-degenerate eigenmode (non-degenerate subradiance),  as in other systems \cite{Cabot_PRL}. Moreover, we find that for both mechanisms, synchronization is related to interference effects in the correlation spectrum that yield a fine structure in a frequency range smaller than the scale fixed by the rates of the intrinsic incoherent processes.

In this work, we have  mainly focused our investigation on a specific model of two qubits, both relevant and analytically treatable, in which we have been able to analyze and compare these phenomena in detail. However, we remark that the phenomenology here presented can arise in different physical scenarios, for which we have provided specific examples. In particular, we have shown that synchronization due to either coalescence or non-degenerate subradiance can emerge in larger systems and even with different kinds of interactions, such as local dissipation and coherent coupling. Hence, we expect that our detailed analysis might help to the observation of synchronization and the recognition of the mechanisms enabling it in more general contexts.

\section*{Acknowledgments}

The authors acknowledge support from MINECO/AEI/FEDER through projects EPheQuCS FIS2016-78010-P, CSIC Research Platform PTI-001,  the QUAREC project funded by CAIB, the María de Maeztu Program for Units of Excellence in R\&D (MDM-2017-0711), and funding from CAIB PhD and postdoctoral programs. 

\appendix

\section{Liouville formalism}\label{appA}

\subsection{Liouville representation of the master equation}

The master equation (\ref{ME}) describing the evolution of $\hat{\rho}$ can be rewritten as $\dot{\hat{\rho}}=\mathcal{L}\hat{\rho}$, where $\mathcal{L}$ is the Liouvillian superoperator. In the Liouville representation, the state of the system is represented by a vector of the Hilbert-Schmidt space $\mathcal{H}=\mathbb{C}^{16}$ and $\mathcal{L}$ is a non-Hermitian matrix (more details can be found in Refs. \cite{Bellomo,Marco2}). The vector in the Hilbert-Schmidt space representing the state of the system is  $|\rho\rrangle$, which is obtained through a mapping that corresponds to a row-major vectorization\footnote{This kind of vectorization mapping, $\text{vec}(\cdot)$, transforms the density matrix $\hat{\rho}$  to a column vector $|\rho\rrangle=\text{vec}(\hat{\rho})$ by arranging consecutively its rows, while a product of operators transforms as $\text{vec}(\hat{o}_1\hat{\rho}\hat{o}_2)=(\hat{o}_1\otimes \hat{o}_2^\top)\text{vec}(\hat{\rho})$.}:
\begin{equation}\label{DefinitionRL}
\hat{\rho} = \sum_{i,j=1}^4 \rho_{ij} |i\rangle \langle j| \rightarrow \Ket{\rho}
= \sum_{i,j=1}^4 \rho_{ij}  \Ket{ij } ,
\end{equation}
with $\Ket{ij}=|i\rangle\otimes|j\rangle$. In this space, vectors are denoted as $\Ket{\cdot}$ while $\Bra{\cdot}$ correspond to their  conjugate transpose partners. The inner product  is defined as $\llangle v_2|v_1\rrangle=\text{Tr}(\hat{v}_2^\dagger \hat{v}_1)$ where $\hat{v}_1(\hat{v}_2^\dagger)$ are the matrices obtained by mapping $|v_1\rrangle(\llangle v_2|)$ back into the Hilbert space. Then, the matrix representation of $\mathcal{L}$  is given by
\begin{widetext}
\begin{equation}\label{m_L}
\begin{split}
\mathcal{L}=-i(\hat{H}\otimes \mathbb{I} -\mathbb{I}\otimes \hat{H}^\top)+\sum_{i,j=1}^2 \gamma \big[\hat{\sigma}_i^-\otimes(\hat{\sigma}_j^+)^\top -(\hat{\sigma}_j^+\hat{\sigma}^-_i)\otimes\frac{\mathbb{I}}{2}-\frac{\mathbb{I}}{2}\otimes(\hat{\sigma}_j^+\hat{\sigma}^-_i)^\top\big]\\
+\sum_{i=1}^2 w \big[\hat{\sigma}_i^+\otimes(\hat{\sigma}_i^-)^\top -(\hat{\sigma}_i^-\hat{\sigma}^+_i)\otimes\frac{\mathbb{I}}{2}-\frac{\mathbb{I}}{2}\otimes(\hat{\sigma}_i^-\hat{\sigma}^+_i)^\top\big].
\end{split}
\end{equation}
\end{widetext}
An important feature for this kind of system is that the Liouvillian matrix takes a block-diagonal form \cite{Bellomo,Marco2}: $\mathcal{L}=\bigoplus_\mu \mathcal{L}_\mu$, with $\mu \in\{a,b,c,d,e\}$.
In the same way the Hilbert-Schmidt space $\mathcal{H}$ can be decomposed in these same blocks or subspaces $\mathcal{H}=\bigoplus_\mu \mathcal{H}_\mu$ each of which is spanned by the following basis elements: subspace $\mathcal{H}_a$ is spanned by $|eeee\rrangle$,  $|egeg\rrangle$,  $|egge\rrangle$,  $|geeg\rrangle$,  $|gege\rrangle$, and  $|gggg\rrangle$; $\mathcal{H}_b$ by  $|eeeg\rrangle$,  $|eege\rrangle$,  $|eggg\rrangle$, and  $|gegg\rrangle $; $\mathcal{H}_c$ by   $|egee\rrangle$,  $|geee\rrangle$,  $|ggeg\rrangle$, and $|ggge\rrangle$; $\mathcal{H}_d$ by  $|eegg\rrangle$; and $\mathcal{H}_e$ by  $|ggee\rrangle$. Then the different Liouvillian blocks read as
\begin{widetext}
%\begin{small}
\begin{equation}\label{La}
\mathcal{L}_a = 
 \begin{pmatrix}
  -2\gamma & w & 0 & 0 & w & 0 \\
  \gamma & -(\gamma+w) & -\frac{\gamma}{2}+is_{12} & -\frac{\gamma}{2}-is_{12} & 0 & w\\
  \gamma & -\frac{\gamma}{2}+is_{12} & -(\gamma+w)-i\delta & 0 & -\frac{\gamma}{2}-is_{12} & 0 \\
  \gamma & -\frac{\gamma}{2}-is_{12} & 0 & -(\gamma+w)+i\delta  & -\frac{\gamma}{2}+is_{12} & 0 \\
\gamma & 0 & -\frac{\gamma}{2}-is_{12} & -\frac{\gamma}{2}+is_{12} & -(\gamma+w) & w\\
0 & \gamma & \gamma & \gamma & \gamma & -2w
 \end{pmatrix},
\end{equation}
%\end{small}
\\
\\

%\begin{small}
\begin{equation}\label{Lb}
\mathcal{L}_b=
\begin{pmatrix}
-\frac{3\gamma+w}{2}-i(\omega_0-\frac{\delta}{2}) & -\frac{\gamma}{2}+is_{12} & 0 & w\\
-\frac{\gamma}{2}+is_{12} & -\frac{3\gamma+w}{2}-i(\omega_0+\frac{\delta}{2}) & w & 0\\
\gamma & \gamma & -\frac{\gamma+3w}{2}-i(\omega_0+\frac{\delta}{2}) & -\frac{\gamma}{2}-is_{12}\\
\gamma & \gamma & -\frac{\gamma}{2}-is_{12} & -\frac{\gamma+3w}{2}-i(\omega_0-\frac{\delta}{2})
\end{pmatrix},
\end{equation}
%\end{small}
\end{widetext}
\noindent $\mathcal{L}_c$ is the complex conjugate of $\mathcal{L}_b$, $\mathcal{L}_d=-(\gamma+w)-2i\omega_0$, and $\mathcal{L}_e=(\mathcal{L}_d)^*$. 

\subsection{Analytical expressions for the eigenvalues}

\begin{figure}[t!]
 \centering
 \includegraphics[width=0.9\columnwidth]{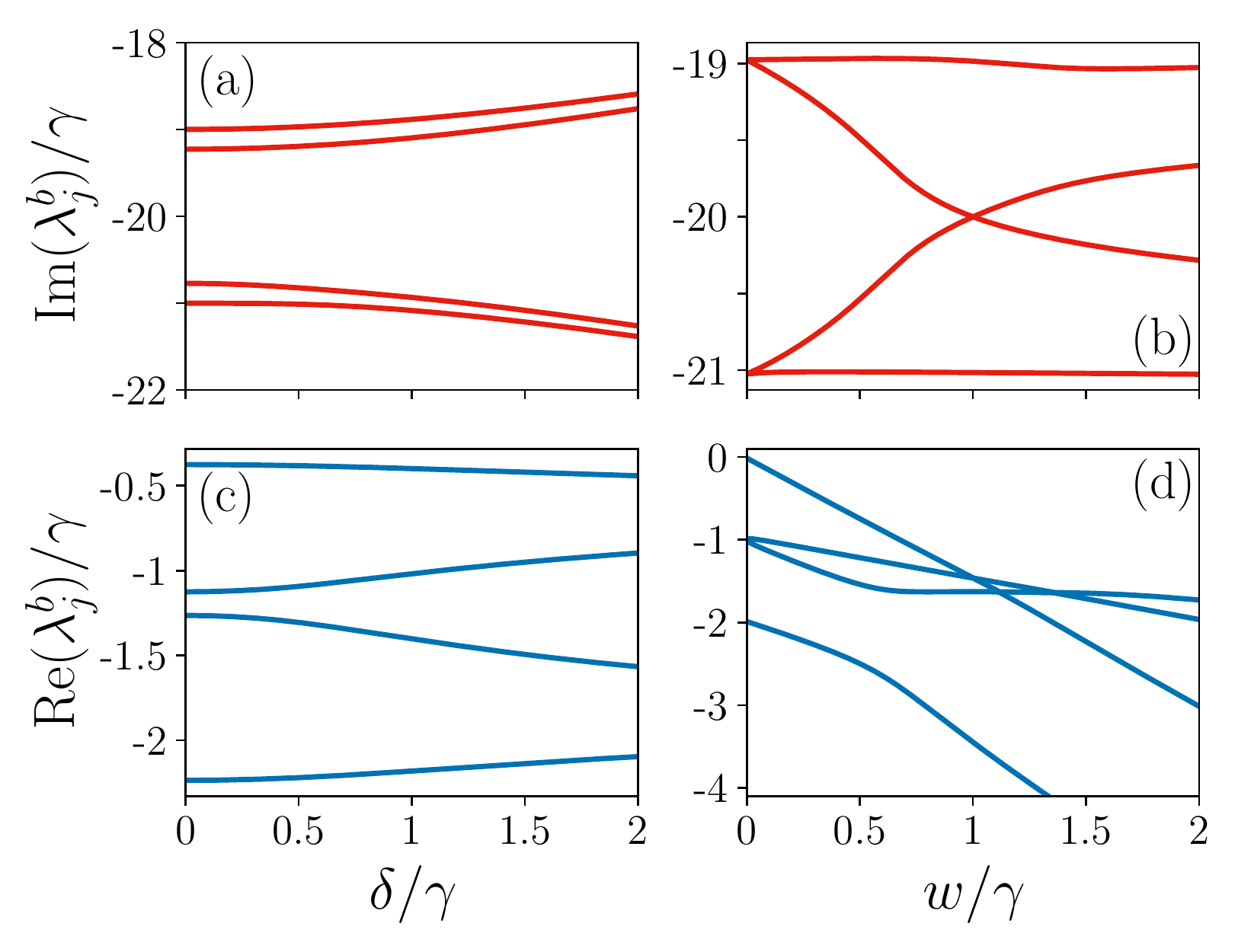}
 \caption{Eigenfrequencies (a,b) and decay rates (c,d) varying $\delta/\gamma$ (a,c), or $w/\gamma$ (b,d). In both cases $\omega_0/\gamma=20$ and $s_{12}/\gamma=1$, while in (a,c) $w/\gamma=0.25$ and in (b,d) $\delta/\gamma=0.5$.}
 \label{eigs_fig}
\end{figure}

In the most general case in which all parameters are nonzero, the analytical expressions for the complete set of eigenvalues of these matrices $\lambda_k^\mu$  are very cumbersome and will not be reported here. Nevertheless, for some particular cases, useful analytical expressions can be found. In fact for $w/\gamma=0$ the full eigenspectrum can be obtained \cite{Bellomo}. The eigenvalues of $\mathcal{L}_b$, which are the relevant ones for our synchronization analysis, are:
\begin{equation}\label{eigs_b1}
\begin{split}
\lambda_1^b&=-\frac{1}{2}[3\gamma+V^*]-i\omega_0,\\
\lambda_2^b&=-\frac{1}{2}[3\gamma-V^*]-i\omega_0,\\
\lambda_3^b&=-\frac{1}{2}[\gamma+V]-i\omega_0,\\
\lambda_4^b&=-\frac{1}{2}[\gamma-V]-i\omega_0,
\end{split}
\end{equation}
ordered with increasing real part and $V=\sqrt{(\gamma+i2s_{12})^2-\delta^2}$. Notice that for $\delta=0$ the real part of $\lambda_4^b$ is zero.  The appearance of purely imaginary eigenvalues corresponds to the existence of decoherence-free subspaces which enable  the possibility of stationary synchronization \cite{manzano,cabot_npj,dieter2}. It is also useful (and possible) to write down the eigenvalues for the case with $\delta/\gamma=0$ and nonvanishing pumping, in which we have:
\begin{equation}\label{eigs_b2}
\begin{split}
\lambda_1^b&=-\frac{1}{2}[3\gamma+2w+\tilde{V}]-i\omega_0,\\
\lambda_2^b&=-\frac{1}{2}[3\gamma+2w-\tilde{V}]-i\omega_0,\\
\lambda_3^b&=-\gamma-\frac{w}{2}-i(\omega_0+s_{12}),\\
\lambda_4^b&=-\frac{3}{2}w-i(\omega_0-s_{12}),
\end{split}
\end{equation}
with $\tilde{V}=\sqrt{(w^2+\gamma^2+6w\gamma-4s_{12}^2)+i4s_{12}(w-\gamma)}$. Here we can find two EPs,  one for $s_{12}=0$ and $w/\gamma=1$ in which   $\lambda_3^b=\lambda_4^b$ and their respective eigenvectors coalesce, and the other at $s_{12}/\gamma=\sqrt{2}$ and $w/\gamma=1$ in which the ones coalescing are  $\lambda_2^b=\lambda_1^b$. The behavior of the EPs for $w/\gamma=1$ is shown in Fig. \ref{EP_s12} in which, as mentioned in the main text, varying the coupling and the detuning up to three EPs appear. Finally notice that for $s_{12}=0$ and  $w/\gamma=2/3$, we have $\lambda_4^b=\lambda_2^b$, but this kind of degeneracy is a trivial one and does not bring any coalescence, as can be seen looking at the eigenvector multiplicity across this point. In Fig. \ref{eigs_fig} we show the typical eigenvalue trajectory in the absence of coalescence, and varying different parameters of the system. We highlight how the branching behavior of Fig. \ref{F123} disappears in the absence of EPs.

\subsection{EPs in $\mathcal{L}_a$}

\begin{figure}[t!]
 \centering
 \includegraphics[width=0.9\columnwidth]{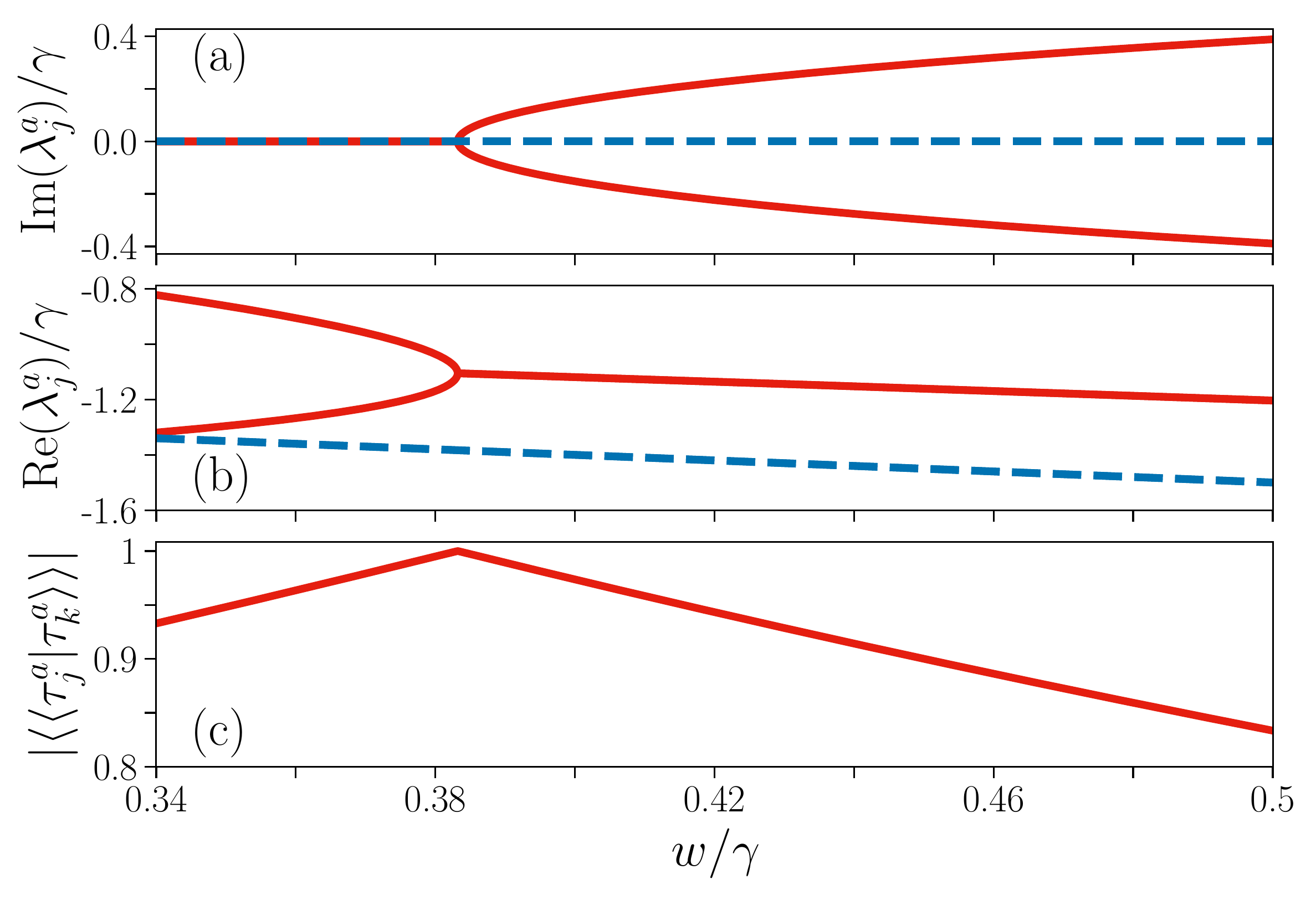}
 \caption{(a) Imaginary part of the eigenvalues (eigenfrequencies) of $\mathcal{L}_a$, varying $w/\gamma$, for $\delta/\gamma=0$, $s_{12}/\gamma=1$ and $\omega_0/\gamma=20$. In solid red the pair of eigenvalues that coalesce. (b) The real part of the corresponding eigenvalues (decay rates). (c) Product of the corresponding pair of eigenvectors that coalesce. Notice that not all eigenvalues are visible, as we have adjusted the range of the plots to display clearly  the EP.}
 \label{EpLa}
\end{figure}

In this section we show an example of EP in $\mathcal{L}_a$. In this sector and for the case $w\neq0$ and $\delta=0$ there are three eigenvalues with simple expressions:
\begin{equation}\label{eigs_a1}
\begin{split}
\lambda^a_1&=0,\\
\lambda^a_2&=-(w+\gamma)-2is_{12},\\
\lambda^a_3&=-(w+\gamma)+2is_{12},
\end{split}
\end{equation}
while the remaining three are roots of  the third order equation:
\begin{equation}\label{eigs_a2}
\begin{split}
\lambda^3+4\lambda^2(w+\gamma)+\lambda(5w^2+10w\gamma+4\gamma^2)\\
+2w^3+6w^2\gamma+8w\gamma^2=0.
\end{split}
\end{equation}
Notice that here the eigenvalues are not ordered. Without the need of finding the solutions of Eq. (\ref{eigs_a2}) we can readily obtain important information. First notice that for $w=0$ there is a second eigenvalue together with $\lambda^a_1$ which is zero, and thus the stationary state is not unique. In fact for $\delta=w=0$ we have shown that there are pure imaginary eigenvalues in $\mathcal{L}_{b(c)}$, which represent the non-decaying oscillating coherences between the two steady states, which attain the possibility of stationary synchronization \cite{manzano,cabot_npj,dieter1,dieter2}. Second, notice that as a third order equation can have either three real roots or one real root and two complex conjugate ones, the corresponding branching of eigenvalues resembles what has been discussed for $\mathcal{L}_{b(c)}$ and thus there could be an EP at the branching point. This turns out to be the case, as we show in Fig. \ref{EpLa} in which at the point in which two roots become complex, the corresponding eigenvectors become parallel.

\subsection{Dynamics of $\langle \hat{\sigma}_j^x(t)\rangle$}

Here we write down the formal solution for the dynamics of $\langle \hat{\sigma}_j^x(t)\rangle$ in terms of coefficients that depend on the eigenvalues and eigenvectors of $\mathcal{L}_{b(c)}$. Notice that as it depends on the diagonalization of $\mathcal{L}$, it is not valid at an EP (see for instance Ref. \cite{Longhi1}). Moreover, as the analytical expressions for the eigenspectrum of the system are in general too cumbersome, the following solution is usually complemented by the numerical calculation of its coefficients. The semi-analytical solution is obtained proceeding as follows \cite{Bellomo}. We first notice that the density matrix at any time can be written as\footnote{The identity in the Hilbert-Schmidt space can be written as $\mathcal{I}=\bigoplus_\mu\sum_k\frac{|\tau_k^\mu\rrangle\llangle \bar{\tau}_k^\mu|}{\llangle \bar{\tau}^\mu_k|\tau^\mu_k\rrangle}$ when $\mathcal{L}$ is diagonalizable.}
\begin{equation}
|\rho(t)\rrangle=\sum_{\mu}\sum_{k} p^\mu_{0k}|\tau_k^\mu\rrangle e^{\lambda_k^\mu t} 
\end{equation}
where the initial condition is encoded in the coefficients $p^\mu_{0k}$ with $\mu \in\{a,b,c,d,e\}$, defined as the overlap of $\hat{\rho}(0)$ with the right (left) eigenvectors of the Liouvillian $|\tau^\mu_k(\bar{\tau}^\mu_k)\rrangle$, i.e.  $p^\mu_{0k}=\llangle \bar{\tau}^\mu_k|\rho(0)\rrangle/\llangle \bar{\tau}^\mu_k|\tau^\mu_k\rrangle$. Then from the definition of expected value we obtain
\begin{equation}
\langle \hat{\sigma}_j^x(t)\rangle=\text{Tr}(\hat{\sigma}_j^x\hat{\rho}(t))=\sum_{\mu}\sum_{k} p^\mu_{0k} \langle\tau_k^\mu\rangle_{xj}e^{\lambda_k^\mu t}, 
\end{equation}
with $\langle\tau_k^\mu\rangle_{xj}=\llangle \sigma_j^x|\tau_k^\mu\rrangle$ and, invoking the block structure of the Liouvillian, we find that $\langle\tau_k^\mu\rangle_{xj}$ are nonzero only for $\mu=b,c$. Finally, as $\mathcal{L}_c=\mathcal{L}_b^*$, then $\lambda^c_k=\lambda^{b*}_k$, $\langle\tau_k^c\rangle_{xj}=\langle\tau_k^b\rangle_{xj}^*$ and  $p^c_{0k}=p^{b*}_{0k}$. Thus the formal solution can be written just in terms of $\mu=b$ as
%
%\small
\begin{equation}\label{solution}
\langle \hat{\sigma}_j^x(t)\rangle=\sum_{k=1}^4 2|p^b_{0k}\langle\tau_k^b\rangle_{xj}|e^{\text{Re}(\lambda_k^b)t}\cos[\text{Im}(\lambda_k^b)t+\psi_{k,xj}^b], 
\end{equation}
%\normalsize
%
with $\psi_{k,xj}^b=\text{arg}(p^b_{0k}\langle\tau_k^b\rangle_{xj})$.

\section{Synchronization measure}\label{appB}

In this section we present the measure that we use to assess the presence of synchronization, which consists in a correlation function that quantifies the degree of similitude between two temporal trajectories \cite{SyncRev1,SyncRev2}. In particular, these trajectories correspond to local observables of each system, as for instance $A_1(t)=\langle \hat{\sigma}^x_1(t)\rangle$ and $A_2(t)=\langle \hat{\sigma}^x_2(t)\rangle$, for some particular parameter choice and initial condition. The corresponding correlator is the Pearson factor defined as:
\small
\begin{equation}\label{SyncMeasure}
\mathcal{C}_{A_1(t),A_2(t)}(\Delta t)=\frac{\int_t^{t+\Delta t}ds[A_1(s)-\bar{A}_1][A_2(s)-\bar{A}_2] }{\sqrt{\prod_{j=1}^2 \int_t^{t+\Delta t}ds[A_j(s)-\bar{A}_j]^2}},
\end{equation}
\normalsize
\noindent with $\bar{A}_j=\frac{1}{\Delta t}\int_t^{t+\Delta t}ds A_j(s)$. Then $\mathcal{C}_{A_1(t),A_2(t)}(\Delta t)\in[-1,1]$ by definition. This correlator is a function of time with a time window $\Delta t$, which for perfect synchronization or anti-phase synchronization  is known to take the values 1 or -1, respectively. However, an important drawback is that it is not sensitive to synchronization at other phase-differences. For this reason, and in order to assess the emergence of synchronization with arbitrary {\it locked} phase differences, we consider the time delayed maximized Pearson factor. This is defined as $\mathcal{C}_{\text{max}}=\text{max}\big[\mathcal{C}_{A_1(t),A_2(t+\tau)}(\Delta t) \big]_{\tau\in[0,\delta t]}$, or in words: it is the maximum value that the Pearson factor takes considering two time delayed trajectories with a delay time in the range 0 to $\delta t$. This measure takes the value 1 for perfect synchronization. Notice that in this case, from the optimal $\tau$ we can obtain the locked phase difference between the synchronized trajectories. At this point we should remark that there are not universal prescribed values for $\delta t$ and $\Delta t$, rather there is a qualitative recipe for them to be meaningful: $\delta t$ should be of the order of a period of the synchronous oscillation, and $\Delta t$ should be of the order of few periods of the synchronous oscillation. 

\section{Correlation spectrum for $w/\gamma=0$}\label{appC}

In this section we outline  the main steps involved in computing two-time correlations of the type $\langle\hat{\sigma}_{j}^-(t+\tau)\hat{\sigma}^+_{k}(t)\rangle$ in the stationary state of the system, that is $\langle\hat{\sigma}_{j}^-(\tau)\hat{\sigma}^+_{k}(0)\rangle_{ss}=\text{lim}_{t\to\infty}\langle\hat{\sigma}_{j}^-(t+\tau)\hat{\sigma}^+_{k}(t)\rangle$. In the absence of pumping the stationary state of the system is the vacuum, $\rho_{ss}=|gg\rangle\langle gg|$. Using the quantum regression theorem \cite{Carmichael} we have
\begin{eqnarray}
\langle \hat{\sigma}_j^-(\tau)\hat{\sigma}_k^+(0)\rangle_{ss}&=\text{Tr}\big( \hat{\sigma}_j^-e^{\mathcal{L}\tau}(\hat{\sigma}_k^+|gg\rangle\langle gg|)\big)\nonumber\\
&=\text{Tr}\big( \hat{\sigma}_j^-e^{\mathcal{L}_b\tau}(\hat{\sigma}_k^+|gg\rangle\langle gg|)\big),
\end{eqnarray}
for $\tau\geq0$. In the second equality we have used the fact that $\hat{\sigma}_k^+|gg\rangle\langle gg|$ yields either $|eg\rangle\langle gg|$ or $|ge\rangle\langle gg|$ whose dynamics is ruled by $\mathcal{L}_b$. Moreover, as $w/\gamma=0$, and as this type of initial condition belongs to the one excitation sector, the dynamics of these correlations can be obtained just considering the one excitation sector. Thus, considering a more general initial condition of this type, we have that $e^{\mathcal{L}_b\tau}(\rho_{eggg}(0)|eg \rangle\langle gg|+\rho_{gegg}(0)|ge \rangle\langle gg|)=\rho_{eggg}(\tau)|eg \rangle\langle gg|+\rho_{gegg}(\tau)|ge \rangle\langle gg|$, where these amplitudes follow a system of equations given by $\mathcal{L}_b$ that reads as
\begin{widetext}
%
%\small
\begin{equation}
\begin{split}
\partial_\tau  \rho_{eggg}(\tau)=-\big[\frac{\gamma}{2}+i(\omega_0+\frac{\delta}{2})\big]\rho_{eggg}(\tau)-(\frac{\gamma}{2}+is_{12})\rho_{gegg}(\tau),\\
\partial_\tau  \rho_{gegg}(\tau)=-\big[\frac{\gamma}{2}+i(\omega_0-\frac{\delta}{2})\big]\rho_{gegg}(\tau)-(\frac{\gamma}{2}+is_{12})\rho_{eggg}(\tau).
\end{split}
\end{equation}
%\normalsize
%
The solution in the Laplace domain, $\rho_{xxgg}(s)=\int_0^\infty \rho_{xxgg}(\tau)e^{-s\tau}d\tau$, is readily obtained
%
%\small
\begin{equation}\label{laplace_sol}
\begin{split}
\rho_{eggg}(s)=\frac{[s+\gamma/2+i(\omega_0-\delta/2)]\rho_{eggg}(0)-(\gamma/2+is_{12})\rho_{gegg}(0)}{(s-\lambda_3^b)(s-\lambda_4^b)},\\
\rho_{gegg}(s)=\frac{[s+\gamma/2+i(\omega_0+\delta/2)]\rho_{gegg}(0)-(\gamma/2+is_{12})\rho_{eggg}(0)}{(s-\lambda_3^b)(s-\lambda_4^b)},
\end{split}
\end{equation}
%\normalsize
%
\end{widetext}
where the poles correspond to two of the eigenvalues given in Eq. (\ref{eigs_b1}). Notice that for $s_{12}=0$ there is an EP at $\delta=\gamma$ but, in contrast to Eq. (\ref{solution}), this solution is correct at the EP as it is not written in terms of the eigenvectors of $\mathcal{L}_b$. Moreover, the EP appears as a double pole, with the direct consequence of an anomalous decay dynamics at this point, in which the exponentials present polynomial corrections in time (see also \cite{Cabot_EPL}). We can consider collective measurements or individual ones, each case corresponding to different linear combinations of the above general results. For instance, for the collective correlation function associated to  $\hat{L}=(\hat{\sigma}_1^-+\hat{\sigma}_2^-)/\sqrt{2}$, we have $\langle \hat{L}(\tau)\hat{L}^\dagger(0)\rangle_{ss}=(\rho_{eggg}(\tau)+\rho_{gegg}(\tau))/\sqrt{2}$ with the initial condition $\rho_{eggg}(0)=1/\sqrt{2}$ and $\rho_{gegg}(0)=1/\sqrt{2}$. Otherwise, considering only the initial excitation of one of the qubits, we have $\langle\hat{\sigma}_{1}^-(\tau)\hat{\sigma}^+_{1}(0)\rangle_{ss}=\rho_{eggg}(\tau)$ and  $\langle\hat{\sigma}_{2}^-(\tau)\hat{\sigma}^+_{2}(0)\rangle_{ss}=\rho_{gegg}(\tau)$ for either $\rho_{eggg}(0)=1$ and $\rho_{gegg}(0)=0$ or the other way around.

In general we will be interested in the Fourier transform or spectrum of these correlations, i.e.
\begin{eqnarray}\label{out_spec}
\mathcal{S}_{\hat{o}\hat{o}^\dagger}(\omega)&=&\int_{-\infty}^\infty d\tau \,e^{-i\omega \tau} \langle\hat{o}(\tau)\hat{o}^\dagger\rangle_{ss}\nonumber\\
&=&2\text{Re}\bigg\{\int_{0}^\infty d\tau \,e^{-i\omega \tau} \langle\hat{o}(\tau)\hat{o}^\dagger)\rangle_{ss} \bigg\},
\end{eqnarray}
 where $\hat{o}$ stands either for $\hat{\sigma}_j^-$ or $\hat{L}$. The second equality in (\ref{out_spec}) follows from the fact that in the stationary state $\langle\hat{o}(-\tau)\hat{o}^\dagger\rangle_{ss}= \langle\hat{o}\hat{o}^\dagger(\tau)\rangle_{ss}$, and moreover for these correlations $\langle\hat{o}\hat{o}^\dagger(\tau)\rangle_{ss}=\langle\hat{o}(\tau)\hat{o}^\dagger\rangle_{ss}^*$. Finally notice that  these Fourier transformed correlations can be written in terms of the solutions in the Laplace domain as combinations of the terms $2\text{Re}[\rho_{eggg}(s=i\omega)]$ and $2\text{Re}[\rho_{gegg}(s=i\omega)]$.

\section{Details on the 1D arrays of qubits}\label{appd}

\subsection{Array with dissipative couplings}

In the one-excitation sector the dynamics of the coherences described by the master equation (\ref{ME3}), is given by the following system of equations:

\small
\begin{eqnarray}
\partial_t\langle \hat{\sigma}^-_{Aj}\rangle=-(i\omega_1+\gamma) \langle \hat{\sigma}^-_{Aj}\rangle -\frac{\gamma}{2}\big(\langle\hat{\sigma}^-_{B(j-1)}\rangle(1-\delta_{j,1})+\langle\hat{\sigma}^-_{Bj}\rangle\big),\nonumber\\
\partial_t\langle\hat{\sigma}^-_{Bj}\rangle=-(i\omega_2+\gamma) \langle \hat{\sigma}^-_{Bj}\rangle -\frac{\gamma}{2}\big(\langle\hat{\sigma}^-_{Aj}\rangle+\langle\hat{\sigma}^-_{A(j+1)}\rangle(1-\delta_{j,N})\big),\nonumber
\end{eqnarray}
\normalsize

\noindent with $j\in[1,N]$, and $\delta_{j,j'}$ the Kronecker delta. We can obtain the eigenvalues of this system performing the following orthogonal transformation \cite{Cabot_PRL,Cabot_EPL}:

\begin{equation}\label{akl}
   \langle \hat{\sigma}^-_{A_{k_l}}\rangle=\sum_{j=1}^N \mathcal{S}^{(A)}_{j,k_l}\langle \hat{\sigma}^-_{Aj}\rangle ,\quad 
\langle \hat{\sigma}^-_{Aj}\rangle=\sum_{l=1}^N \mathcal{S}^{(A)}_{j,k_l} \langle\hat{\sigma}^-_{A_{k_l}}\rangle ,
\end{equation}
\begin{equation}\label{bkl}
   \langle \hat{\sigma}^-_{B_{k_l}}\rangle=\sum_{j=1}^N \mathcal{S}^{(B)}_{j,k_l}\langle \hat{\sigma}^-_{Bj}\rangle ,\quad 
\langle \hat{\sigma}^-_{Bj}\rangle=\sum_{l=1}^N \mathcal{S}^{(B)}_{j,k_l} \langle\hat{\sigma}^-_{B_{k_l}}\rangle ,
\end{equation}
where the mode functions are defined as
\small
\begin{equation}\nonumber
\mathcal{S}^{(A)}_{j,k_l}=\sqrt{\frac{2}{N+\frac{1}{2}}}\sin[k_l(j-\frac{1}{2})],\quad \mathcal{S}^{(B)}_{j,k_l}=\sqrt{\frac{2}{N+\frac{1}{2}}}\sin(k_lj),
\end{equation}
\normalsize
with $k_l={\pi l}/{(N+1/2)}, \quad l=1,\dots,N$. This defines the orthogonal transformation which satisfies $\sum_{j=1}^N \mathcal{S}^{(x)}_{j,k_l}\mathcal{S}^{(x)}_{j,k_{l'}}=\delta_{l,l'}$ and $\sum_{l=1}^N \mathcal{S}^{(x)}_{j,k_l}\mathcal{S}^{(x)}_{j',k_{l}}=\delta_{j,j'}$ ($x=A,B$). After this transformation the system of equations for the coherences in the one excitation sector reduces to a block diagonal form, made of $N$ two dimensional blocks, from which we obtain the eigenvalues given in Eq. (\ref{lambdak}). The $N$ blocks in $k$-space read as
\begin{eqnarray}
\partial_t\langle \hat{\sigma}^-_{A_{k_l}}\rangle=-(i\omega_1+\gamma) \langle \hat{\sigma}^-_{A_{k_l}}\rangle -\gamma\cos(k_l/2)\langle\hat{\sigma}^-_{B_{k_l}}\rangle,\nonumber\\
\partial_t\langle\hat{\sigma}^-_{B_{k_l}}\rangle=-(i\omega_2+\gamma) \langle \hat{\sigma}^-_{B_{k_l}}\rangle -\gamma\cos(k_l/2)\langle\hat{\sigma}^-_{A_{k_l}}\rangle.\nonumber
\end{eqnarray}

\subsection{Array with coherent couplings and local losses}

In this second example we also restrict our analysis to the one-excitation sector, for which the equations of the coherences read as
\small
\begin{eqnarray}
\partial_t\langle \hat{\sigma}^-_{Aj}\rangle=-(i\omega_0+\frac{\gamma_A}{2}) \langle \hat{\sigma}^-_{Aj}\rangle -is_{AB}\big(\langle\hat{\sigma}^-_{B(j-1)}\rangle(1-\delta_{j,1})+\langle\hat{\sigma}^-_{Bj}\rangle\big),\nonumber\\
\partial_t\langle\hat{\sigma}^-_{Bj}\rangle=-(i\omega_0+\frac{\gamma_B}{2}) \langle \hat{\sigma}^-_{Bj}\rangle -is_{AB}\big(\langle\hat{\sigma}^-_{Aj}\rangle+\langle\hat{\sigma}^-_{A(j+1)}\rangle(1-\delta_{j,N})\big),\nonumber
\end{eqnarray}
\normalsize

\noindent with $j\in[1,N]$. The eigenvalues of this system can be obtained following the same procedure as before. In this case, after the transformation (\ref{akl})-(\ref{bkl}), the blocks in $k$-space read:
\begin{eqnarray}
\partial_t\langle \hat{\sigma}^-_{A_{k_l}}\rangle=-(i\omega_0+\frac{\gamma_A}{2}) \langle \hat{\sigma}^-_{A_{k_l}}\rangle -i2s_{AB}\cos(k_l/2)\langle\hat{\sigma}^-_{B_{k_l}}\rangle,\nonumber\\
\partial_t\langle\hat{\sigma}^-_{B_{k_l}}\rangle=-(i\omega_0+\frac{\gamma_B}{2}) \langle \hat{\sigma}^-_{B_{k_l}}\rangle -i2s_{AB}\cos(k_l/2)\langle\hat{\sigma}^-_{A_{k_l}}\rangle,\nonumber
\end{eqnarray}
from which we can obtain the eigenvalues as given in  Eq. (\ref{lambdak2}) of the main text. 

Finally, we recall that while we have focused on open boundary conditions, the same expressions for the eigenvalues, Eq. (\ref{lambdak}) and Eq. (\ref{lambdak2}), can be  found in the properly generalized periodic boundary conditions case of each model. The difference when changing the boundary conditions resides in the definition of $k_l$ and of the mode functions $\mathcal{S}^{(x)}_{j,k_{l}}$, and it does not prevent the presence of coalescence.

\end{document}